\documentclass[pre,aps,twocolumn,showpacs,preprintnumbers,superscriptaddress]{revtex4}

\usepackage{graphicx}
\usepackage{dcolumn}
\usepackage{bm}
\usepackage{amssymb}
\usepackage{pifont}
\usepackage{amsmath}
\usepackage{times}
\usepackage[nooneline]{subfigure}

\usepackage{graphicx,psfrag}
\usepackage{xcolor}

\usepackage{soul}

\begin{document}

\title{An exact firing rate model reveals the differential effects 
of chemical versus electrical synapses in spiking networks}

\author{Bastian Pietras}
\affiliation{Faculty of Behavioural and Movement Sciences,  
Amsterdam Movement Sciences \& Institute of Brain and Behavior Amsterdam,
Vrije Universiteit Amsterdam, van der Boechorststraat 9, Amsterdam 1081 BT, The Netherlands.}
\affiliation{Department of Physics, Lancaster University, Lancaster LA1 4YB, United Kingdom.}
\affiliation{Institute of Mathematics, Technical University Berlin, 10623 Berlin, Germany.}
\affiliation{Bernstein Center for Computational Neuroscience Berlin, 10115 Berlin, Germany.}

\author{Federico Devalle}
\affiliation{Department of Information and Communication Technologies,
 Universitat Pompeu Fabra, 08003 Barcelona, Spain.}
\affiliation{Department of Physics, Lancaster University, Lancaster LA1 4YB, United Kingdom.}

\author{Alex Roxin}
\affiliation{Centre de Recerca Matem\`atica, Campus de Bellaterra, Edifici C,  08193
Bellaterra (Barcelona), Spain.}
\affiliation{Barcelona Graduate School of Mathematics, Barcelona, Spain.}

\author{Andreas Daffertshofer}
\affiliation{Faculty of Behavioural and Movement Sciences,  
Amsterdam Movement Sciences \& Institute of Brain and Behavior Amsterdam,
Vrije Universiteit Amsterdam, van der Boechorststraat 9, Amsterdam 1081 BT, The Netherlands.}

\author{Ernest Montbri\'o}
\affiliation{Department of Information and Communication Technologies, 
Universitat Pompeu Fabra, 08003 Barcelona, Spain.}

\date{\today}

\begin{abstract}
Chemical and electrical synapses shape the dynamics of neuronal networks. 
Numerous theoretical studies have investigated how
each of these types of synapses contributes to 
the generation of neuronal oscillations, but their combined effect is less understood. 
This limitation is further magnified by the impossibility of traditional
neuronal mean-field models ---also known as firing rate models, 
or firing rate equations--- to account for electrical synapses. 
Here we introduce a novel firing rate model that exactly describes
the mean field dynamics of
heterogeneous populations of quadratic integrate-and-fire (QIF) neurons 
with both chemical and electrical synapses. 
The mathematical analysis of the firing rate model reveals a well-established 
bifurcation scenario for networks with chemical synapses, 
characterized by a codimension-2 Cusp point and
persistent states for strong recurrent excitatory coupling.
The inclusion of electrical coupling 
generally implies neuronal synchrony by virtue of a supercritical Hopf bifurcation. 
This transforms the Cusp scenario into a bifurcation scenario 
characterized by three codimension-2 points 
(Cusp, Takens-Bogdanov, and Saddle-Node Separatrix Loop),
which greatly reduces 
the possibility for persistent states. 
This is generic for heterogeneous QIF networks with both 
chemical and electrical coupling. Our results agree 
with several numerical studies on the 
dynamics of large networks of
heterogeneous spiking neurons with electrical and chemical coupling. 
\end{abstract}
\pacs{05.45.Xt} 
\maketitle

\section{Introduction}

Collective oscillations and synchrony are prominent features of neuronal circuits,
and are fundamental for the well-timed coordination of neuronal activity.
Such oscillations are profoundly shaped by the presence of chemical 
synapses~\cite{Wan10}. An 
increasing number of experimental studies indicate both the prevalence and 
functional importance of electrical 
synapses (formed by \emph{gap junctions} between neurons) 
in many diverse regions of central nervous systems, 
especially in inhibitory interneurons~\cite{NPR18,Con17,TWG+18}. 
Electrical synapses
participate in mediating synchronization of neuronal network 
activity~\cite{BZ04,CL04}, suggesting that electrical interaction may be 
interrelated with the generation of oscillations via chemical transmission.

The mechanisms by which chemical synapses 
mediate large-scale synchronous activity have been extensively investigated,
 see e.g.~\cite{Wan10,WTK+00}. 
However, only a few studies addressed the synchronization 
of large networks in which neurons are not only interacting via 
excitation and/or inhibition, but also 
via electrical synapses
~\cite{KE04,PGM+07,Erm06,VMG16,HK18,GWP12,TC14,LR03,CK00,OBH09,PMG+03,Coo08,MLP+07}. 
This limited theoretical progress for networks of electrically coupled 
neurons, compared to chemically coupled networks, 
is magnified due to the technical challenges 
faced when developing simplified mean field models 
---often called firing rate models, or firing rate equations (FRE)---
for networks involving electrical synapses. 
While firing rate models turned out to be very useful to 
explain key aspects of the dynamics of 
spiking neuron networks with chemical synapses
~\cite{WC72,DA01,ET10,Hop84,MBT08,BLS95,HS98,TSO+00,RSR15,MR13,TPM98,RBH05,RM11}, 
it remains an open question whether there are similar simplified mean field 
theories for networks involving electrical interactions.

Recently, a novel method has been found to exactly 
derive FRE for populations of heterogeneous quadratic 
integrate-and-fire (QIF) neurons with chemical coupling~\cite{MPR15}. 
The method, related to the so-called Ott-Antonsen 
ansatz~\cite{OA08,OA09,OHA11,PR08,MMS09,PR11,PD16},
allows to obtain exact, low-dimensional firing rate equations for 
ensembles of QIF neurons, see also~\cite{LBS13,SLB14,Lai14}. 
The FRE for QIF neurons have been used to investigate 
numerous problems regarding the dynamics of networks of chemically 
coupled QIF neurons~ 
\cite{PM16,RP16,RM16,DMP18,DRM17,ERA+17,RP17,DEG17,BBC17,Lai18,SAM+18,RP18,DT18,DG19,
ASJ+19,BSD+19,CB19,KBF+19,BAC19,BUA+19}.
Remarkably, previous work has also sought to apply this approach to networks 
with both chemical and electrical coupling~\cite{Lai15}. 
However, in~\cite{Lai15}, the electrical coupling has been treated by making use of 
an approximation which renders the resulting FRE analytically intractable.
We build on this previous work and derive a set of FRE 
for networks with chemical and electrical coupling, but without the 
need for any approximation. 
The resulting system is not only analytically tractable but also allows, in a 
unified framework, for
carrying out a complete analysis of the possible dynamics and bifurcations 
of networks with 
mixed chemical and electrical synapses. 
In {\it Appendix B} we show
that our exact FRE are recovered by 	
appropriately relaxing the approximation invoked in~\cite{Lai15}.

The structure of the paper is as follows: In Section~\ref{secII}, we describe the 
spiking neuron network under investigation, and briefly illustrate the 
impact of electrical coupling in the dynamics of two nonidentical QIF neurons. 
In Section~\ref{secIII}, we introduce the FRE corresponding to the 
thermodynamic limit of the QIF network. The detailed derivation is performed in 
{\it Appendix A}.  
In Section~\ref{secIV}, we perform a comparative analysis of the fixed points 
and their bifurcations in networks with electrical coupling vs. 
networks with chemical coupling. Finally, we investigate the 
dynamics of a QIF network with both electrical and chemical synapses
and demonstrate that the presence of electrical coupling critically determines
the bifurcation scenario of the neuronal network. 
Finally, we discuss our results in Section~\ref{secV}.

\section{Quadratic integrate-and-fire neurons with electrical and chemical synapses}
\label{secII}

We consider a large population of globally electrically and chemically-coupled QIF 
neurons, with membrane potentials $\{V_j\}_{j=1,\dots,N}$ and $N\!\gg\!1$. Their 
dynamics reads  
\begin{equation}
\tau \dot V_j= V_j^2 + \eta_j  + g (v - V_j)+ J \tau s,
\label{qif}
\end{equation}
where $\tau$ denotes the cells' common membrane time constant, and parameter 
$\eta_j$ represents an external input current flowing into cell $j$. 
To model the action potential, the continuous dynamics Eq.~\eqref{qif} is supplemented 
by a discrete resetting rule. Here, we assume that if $V_j$ 
reaches infinity, neuron $j$ emits a spike and its membrane potential is reset to minus 
infinity~\footnote{In our numerical simulations (Euler scheme, $\delta t=10^{-4}$), the resetting rule was applied as follows: When $V_j \geq 100$, the membrane voltage is 
held at $V_j$ for a time interval
$\tau/V_j$. Then, a spike is emitted, and the voltage is reset and kept at $-V_j$ for a 
subsequent interval $\tau/V_j$. In Figs.~\ref{Fig4} and~\ref{Fig7},  
to numerically evaluate the mean membrane potential 
$v$, the population average is computed discarding those neurons with $|V_j|\geq 100$.
\label{numerics}}. 
The mean membrane voltage 
$$v = \frac{1}{N}\sum_{k=1}^N V_k,$$
to which all cells are diffusively coupled with strength $g\geq 0$,
mediates the electrical coupling.
The
constant $J$ quantifies the coupling strength of chemical synapses.
The coupling via chemical synapses is mediated by the mean synaptic activation 
function   
\begin{eqnarray}
s(t)=  \frac{1}{N}\sum_{j=1}^N  \frac{1}{\tau_s}\int\limits_{t-\tau_s}^{t} 
\sum_{k} \delta\!\left(t'-t_j^{k}\right) \, dt',
\label{s}
\end{eqnarray} 
where $t_j^k$ denotes the time of the $k$-th spike of the $j$-th neuron, $\delta (t)$ 
is the Dirac delta function, and $\tau_s$ is a synaptic time constant
~\footnote{We used $\tau_s=10^{-2}$ms, and $\tau=10$~ms. The
instantaneous firing rates shown in Fig.~\ref{Fig4} were obtained by binning 
time and counting spikes within a sliding time window of size
$\delta t=2.5\times10^{-2}$~ms.}.
The synaptic weight $J$ can be positive or negative depending on whether the
chemical synapses are excitatory or inhibitory, respectively

In the absence of coupling, $J=g=0$, 
the QIF neurons are either quiescent ($\eta_i<0$),
or oscillatory ($\eta_i>0$) with frequency 
\begin{equation}
f_i=\frac{1}{\tau \pi}\sqrt{\eta_i}.
\label{freq}
\end{equation}
These two dynamical regimes of individual neurons are connected by a saddle-node
on the invariant circle (SNIC) bifurcation, which occurs when $\eta_i=0$, with $f_i=0$.

Electrical coupling tends to equalize the membrane potentials of the
neurons they connect and may favor synchrony.
Yet, if a large fraction of cells in the network is quiescent,
gap junctions may suppress oscillations and neural synchrony. 
Next we illustrate this phenomenon for two nonidentical QIF neurons
that are coupled via a gap junction~\footnote{This phenomenon 
has been termed `Aging transition' in the literature~\cite{DN04,PM06}.}.
The results of this analysis will later be useful
to understand some aspects of the dynamics of a large network of electrically coupled 
QIF neurons.

\subsection{Strong coupling limit of two electrically coupled QIF neurons}
\label{secII_twoneurons}

\begin{figure}[t]
\psfrag{Vpre}[b][b][1.1]{$V_{pre}$}
\psfrag{Vpost}[b][b][1.1]{$V_{post}$}
\psfrag{time}[b][b][1.1]{$t$}
\centerline{\includegraphics[width=80mm,clip=true]{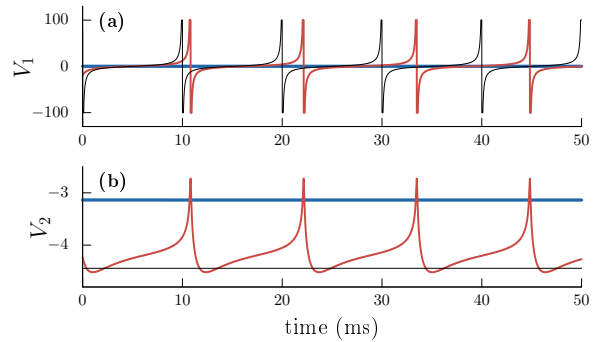}}
\caption{Strong electrical coupling suppresses oscillatory activity for 
negative mean currents, $\bar\eta<0$. 
The panels show the time series of the membrane voltages 
of $N=2$ electrically coupled
QIF neurons. (a) Self-oscillatory 
neuron with $\eta_{1}=\pi^2$; (b) Quiescent neuron with
$\eta_{2}=-2\pi^2$. 
The two neurons are either uncoupled (black thin 
curves, $g=0$), weakly coupled (red curves, $g=1$), 
or strongly coupled (blue thick curves, $g=6$). 
We used $\tau=10$~ms and $J=0$.
}
\label{Fig1}
\end{figure}

We consider a network of $N=2$ nonidentical QIF neurons with dynamics Eq.~\eqref{qif}.
The neurons are coupled via gap junctions only, i.e.~$J=0$ but $g>0$. 
We are interested in the strong coupling limit $g\gg 0$ when $\eta_1>0$ and $\eta_2<0$. 
In Fig.~\ref{Fig1} we depict the corresponding time series of cell 1 
(panel a) and cell 2 (panel b). Black thin curves correspond to the dynamics 
of the uncoupled ($g=0$) cells: cell 1 fires periodically, while cell 2 remains quiescent.
When the neurons are electrically coupled (red curves), 
the membrane voltage of cell 2 displays a series of so-called `spikelets' 
\footnote{ 
Note that these spikelets
depend on the shape of the presynaptic spike, and thus on the particular neuron model 
considered.}.
Moreover, the electrical interaction brings cell 1 closer to its firing threshold 
and, hence, its frequency $f_1$ is reduced. 
When $g$ is increased further, cell 1 becomes quiescent 
(blue thick curves). 

Although analyzing the dynamics of the two cells for
arbitrary coupling strength $g$ is a challenge, there exists a simple and 
general result valid in the large $g$ limit, and of relevance for the large-$N$ analysis 
carried out below. Indeed, for large $g$, the dynamics of the $N=2$ network simply 
depends on the sign of the mean current~\cite{PM06} 
$$\bar \eta=\frac{\eta_1\!+\!\eta_2}{2}.$$ 
For $\bar \eta>0$, the quiescent cell eventually becomes self-oscillatory 
as $g$ is increased from zero. By contrast, for $\bar\eta<0$,
the oscillatory cell eventually turns quiescent in the strong coupling limit; 
see the blue lines in Fig.~\ref{Fig1}~\footnote{The value $\bar\eta=0$ 
determines a boundary separating network oscillations 
and quiescence, cf. Eq~(6) with $p_c^\infty=1/2$ in~\cite{PM06}.}. 
%

\section{Firing Rate Model}
\label{secIII}

In the following, we introduce the FRE corresponding to the thermodynamic 
limit of Eqs.~\eqref{qif}. The detailed derivation of the model closely follows 
the lines of \cite{MPR15} and is given in {\it Appendix A}. 

For $N \to \infty$, one can drop the indices in Eq.~\eqref{qif} 
and define a density function $\rho$ such 
that $\rho(V|\eta,t)~dV$ denotes the fraction of neurons with membrane potentials between 
$V$ and $V\!+\!dV$ and parameter $\eta$ at time $t$.
In the limit of instantaneous synaptic processing, i.e. for $\tau_s\to 0$,  Eq.~\eqref{s} 
reduces to $s(t)= r(t)$ with $r(t)$ being the population-mean firing rate.
If the external currents are distributed according to a Lorentzian 
distribution centered around $\eta=\bar \eta$ with half-width $\Delta$,
\begin{equation}
L_{\Delta,\bar\eta}(\eta)= \frac{1}{\pi} \frac{\Delta}{(\eta-\bar\eta)^2 +\Delta^2},
\label{lorentzian2}
\end{equation}
we find that 
the asymptotic mean-field dynamics evolves according to the following FRE
\footnote{See {\it Appendix B} for the 
comparison of the firing rate model Eqs.\eqref{fre} 
with the FRE derived in~\cite{Lai15}.}
\begin{subequations}
\label{fre}
\begin{eqnarray}
\tau \dot r &=& \tfrac{\Delta}{\tau \pi} + 2  r v - g r, \label{frea}\\
\tau \dot v &=&   v^2 +   \bar \eta - (\pi \tau r)^2 + J \tau r  . \label{freb}
\end{eqnarray}
\end{subequations}
The variables $r$ and $v$ are the mean firing rate and mean 
membrane potential, respectively.
They determine the total voltage density for the network Eq.~\eqref{qif}, which turns
out to be a Lorentzian distribution centered at $v(t)$ and of half-width $\pi \tau r(t)$,
\begin{equation}
\rho(V, t)=\frac{1}{\pi} \frac{\pi \tau r(t)}{\left[V-v(t)\right]^2+\left[\pi \tau r(t)\right]^2 }.
\label{la0}
\end{equation}
The structure of the FRE Eqs.~\eqref{fre} reveals an interesting feature:
Electrical coupling is solely mediated by the firing rate through the negative 
feedback term $-gr$ in the $r$-dynamics Eq.~\eqref{frea}, and not by membrane 
potential differences~
\footnote{This might be understood as follows: The evolution equation for the 
mean membrane potential $v$ is obtained summing up the $N$ differential 
equations Eq.~\eqref{qif}:
$\tfrac{1}{N}\sum_{i=1}^N \dot V_i =\dot v= \tfrac{1}{N}\sum_{i=1}^N (V_i^2 + \eta_i)+Js
+g(v - \tfrac{1}{N}\sum_{i=1}^N V_i)$. Although this is not a closed equation
for $v$ and $r$, one finds that the last term ---corresponding to diffusive coupling--- 
cancels to zero.}.
That is, electrical coupling leads to a narrowing of the voltage distribution Eq.~\eqref{la0}, 
i.e.~a decrease in firing rate. This confirms our initial sketch that electrical coupling 
tends to equalize the neurons' membrane potentials and, under 
suitable conditions, this may promote synchrony. By 
contrast, chemical coupling shifts the center of the distribution Eq.~\eqref{la0} of 
voltages via the feedback term $Jr$ in the $v$-dynamics~Eq.~\eqref{freb}. 
The following phase plane and bifurcation analysis of the 
FRE~\eqref{fre} allows for understanding the collective dynamics of 
the QIF network.

\begin{figure}[t]
\centerline{\includegraphics[width=85mm,clip=true]{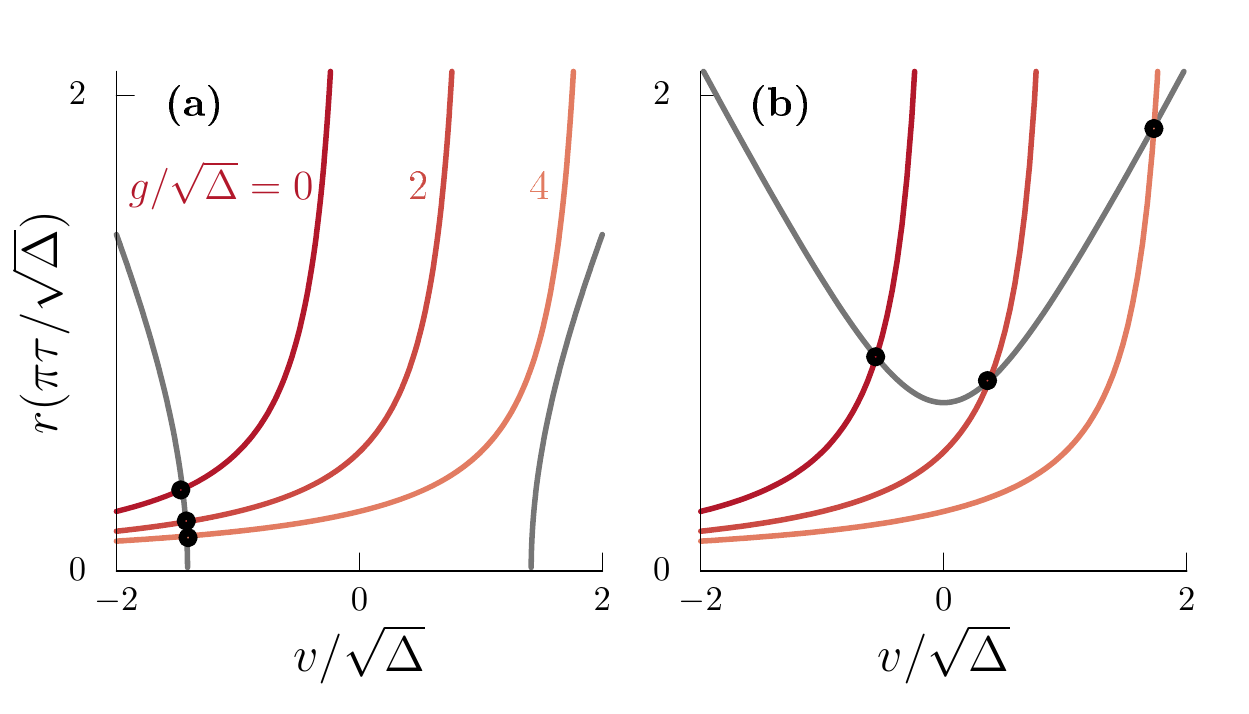}}
\caption{The sign of the mean current $\bar \eta$
determines the behavior of the fixed points of the FRE \eqref{fre} 
with electrical coupling only, $J=0$. The panels show the
nullclines of the FRE~(\ref{fre}) with only electrical coupling ($J=0$) for 
negative ($\bar\eta/\Delta=-2$) and positive ($\bar\eta/\Delta=0.5$) values of $\bar \eta$ 
(panel a and b, respectively), and $g/\sqrt{\Delta}=0,2,4$.
The black points correspond to the intersections of 
$r$-nullclines ($\dot r=0$, red) and $v$-nullclines ($\dot v=0$, gray)
and are fixed points of Eqs.~(\ref{fre}).}
\label{Fig2}
\end{figure}
%

\section{Analysis of the Firing Rate Equations}
\label{secIV}

\subsection{Electrical vs. chemical coupling}

\begin{figure}[t]
\centerline{\includegraphics[width=90mm,clip=true]{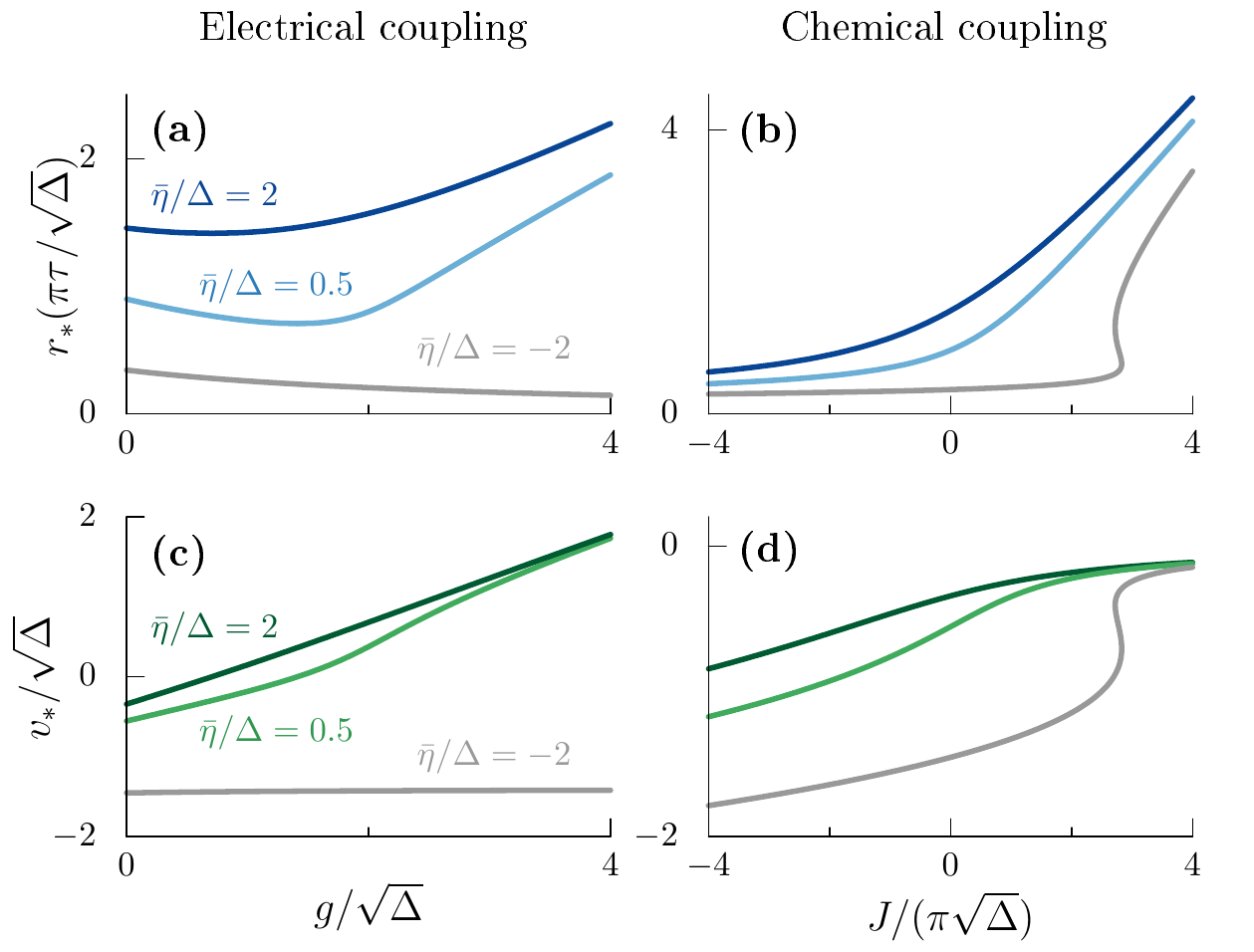}}
\caption{
The bifurcation diagrams of the FRE~(\ref{fre}) for networks with (left) 
electrical and (right) chemical coupling are qualitatively different. 
Top panels 
show the scaled firing rate $r_*/(\pi\tau\sqrt{\Delta})$ versus the ratio of 
(a) electrical $g/\sqrt{\Delta}$ and (b) 
chemical $J/(\pi \sqrt{\Delta})$ synaptic strengths. 
Panels (c,d) show the same bifurcation diagrams for the scaled mean membrane potential 
$v_*/\sqrt{\Delta}$.}
\label{Fig3}
\end{figure}

In the absence of chemical coupling,
our previous discussion of the case $N\!=\!2$ hints at two distinct 
dynamical regimes for positive and negative values of $\bar \eta$. 
With respect to the fixed points $(r_*,v_*)$ of the FRE~\eqref{fre} for $J=0$, 
we find the $v$-nullcline to be 
$$\pi \tau r =\sqrt{v^2+\bar\eta}.$$
Note that if $\bar\eta$ is negative, there exists a range of `forbidden' values of $v$. 
Fig.~\ref{Fig2}(a) shows the nullclines for $\bar\eta<0$ and for
different values of the ratio $g/\sqrt{\Delta}$. Since the majority of 
the neurons are quiescent, an increase in coupling strength $g$ causes active neurons 
to reduce firing, which leads to a progressive decrease of the firing rate $r_*$.
By contrast, in Fig.~\ref{Fig2}(b) the majority of the cells are self-oscillatory, 
$\bar\eta>0$, and strong electrical coupling forces quiescent neurons to 
fire. This yields an increase of $v_*$. Interestingly, the firing rate $r_*$ is a 
non-monotonic function of $g/\sqrt{\Delta}$: 
While $v_*$ remains negative, the voltages are pushed to 
subthreshold values, decreasing the firing rate. This behavior is reverted when $v_*$ 
becomes positive and all voltages are pushed towards values above the firing 
threshold. The different behaviors of Eqs.~\eqref{fre} with electrical coupling for positive 
and negative values of $\bar \eta$ are clearly revealed in the
corresponding bifurcation diagrams shown in Figs.~\ref{Fig3}(a,c).

The case of networks with only chemical coupling, $g=0$, is simpler~\cite{MPR15}. 
The bifurcation diagram depicted in Fig.~\ref{Fig3}(d) shows 
that $v_*$ remains always negative and converges asymptotically to zero as
$J$ increases. The firing rate $r_*$, depicted in Fig.~\ref{Fig3}(b), 
also increases with $J$.
For $\bar \eta<0$ and strong recurrent excitatory coupling, the system undergoes a cusp 
bifurcation and two saddle-node (SN) bifurcations are created. 
This implies the existence of a parameter regime where a persistent, high-activity state 
(stable focus) coexists with a low-activity state (stable node)
---see Fig.~\ref{Fig6}(a), and ~\cite{MPR15}.
This coexistence between persistent and low-activity states  
also occurs in networks with electrical synapses, but it is located 
in a very small region of parameters as we show below, see Fig.~\ref{Fig5}(b).

%

We next explore the linear stability of the fixed points of Eqs.~\eqref{fre},
see also~\emph{Appendix C}. We find that a Hopf bifurcation occurs along the boundary
\begin{equation}
\left(\frac{\bar \eta}{\Delta}\right)_{\text{\scriptsize{H}}}= \frac{4\Delta}{g^2} -
\frac{g^2 }{16\Delta}-\frac{2 J}{\pi  g},
\label{hopf}
\end{equation}
with frequency
\begin{equation}
f_{\text{\scriptsize{H}}}= \frac{1}{\pi \tau }\sqrt{ \bar \eta + \frac{\Delta}{\pi} 
\frac{J }{ g} }.
\label{fH}
\end{equation}
The Hopf boundary Eq.~\eqref{hopf} is depicted in red in the phase diagrams of 
Figs.~\ref{Fig5},~\ref{Fig6}.
Note that $\bar\eta/\Delta\to + \infty$ as $g\to0$ according to Eq.~\eqref{hopf}, 
which indicates that electrical coupling is a necessary ingredient for the Hopf bifurcation 
to exist~
\footnote{For non-instantaneous inhibitory synapses, 
oscillations emerging through a Hopf bifurcation are also 
encountered for weak coupling and weak heterogeneity~\cite{DRM17,DMP18}. 
These oscillations are often referred to as `Interneuronal 
Gamma Oscillations', ING~\cite{WTK+00}. To keep our analysis simple, 
here we do not consider ING oscillations, since Eqs.~\eqref{fre} 
become higher-dimensional and phase plane analysis is no longer possible 
in this case.}.
%

\begin{figure}[t]
\centerline{\includegraphics[width=80mm,clip=true]{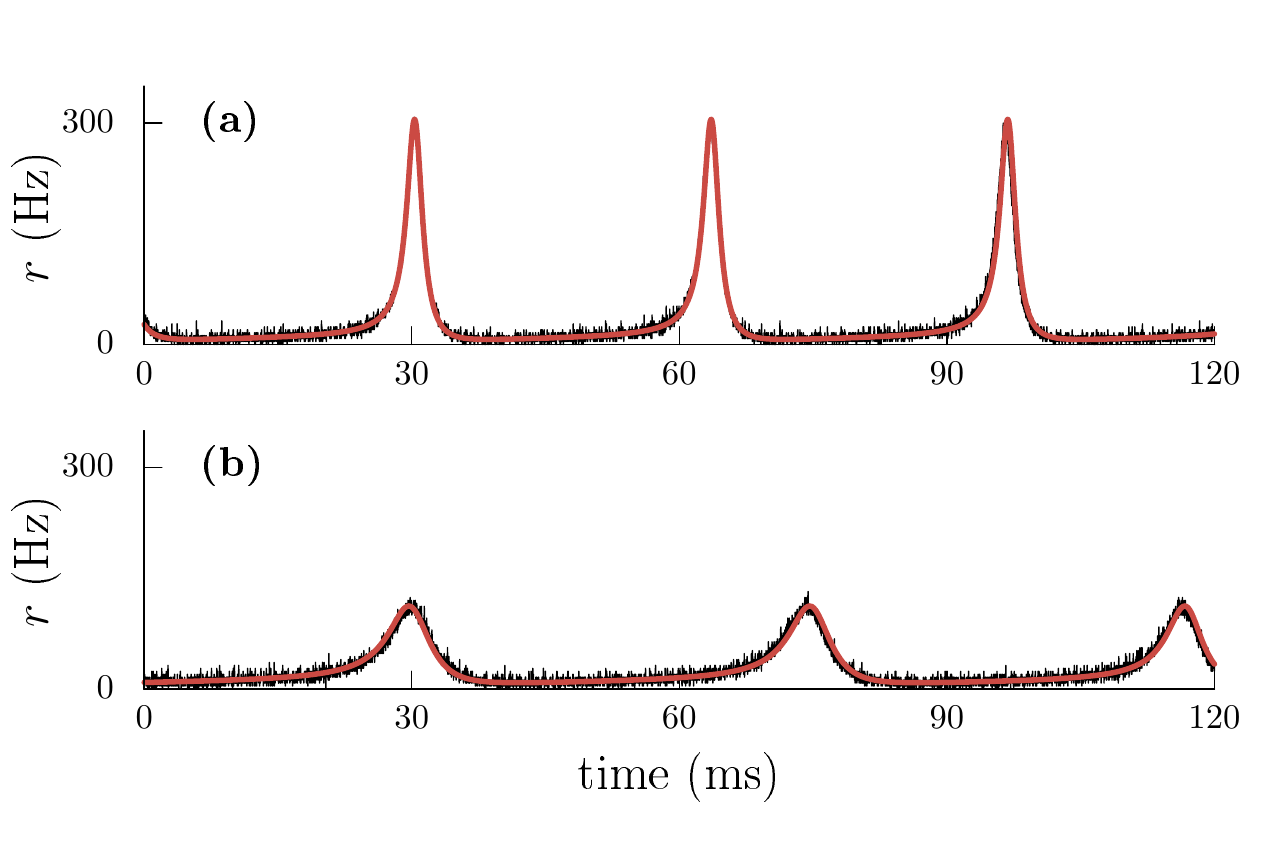}}
\caption{
Electrical coupling promotes collective synchrony. 
The inclusion of inhibitory coupling degrades synchrony and slows down oscillations.
The figure shows the time series of the mean firing rate $r$ of a 
network of $N=10^4$ QIF neurons with dynamics~\eqref{qif} (black)
and of the FRE~\eqref{fre} (red). 
Panel (a) shows collective oscillations (frequency $f\approx 30.1$~Hz)
of a network with gap junctions only ($J=0$). 
Panel (b) corresponds to a network with both 
gap junctions and inhibitory chemical coupling ($J=-\pi$), which both
slows 
down ($f\approx23.6$~Hz) and  
reduces the amplitude of collective oscillations. 
Parameters: $g=3$, $\bar \eta=1$, $\tau=10$~ms and $\Delta=1$.}
\label{Fig4}
\end{figure}

\begin{figure}[t]
\centerline{\includegraphics[width=80mm,clip=true]{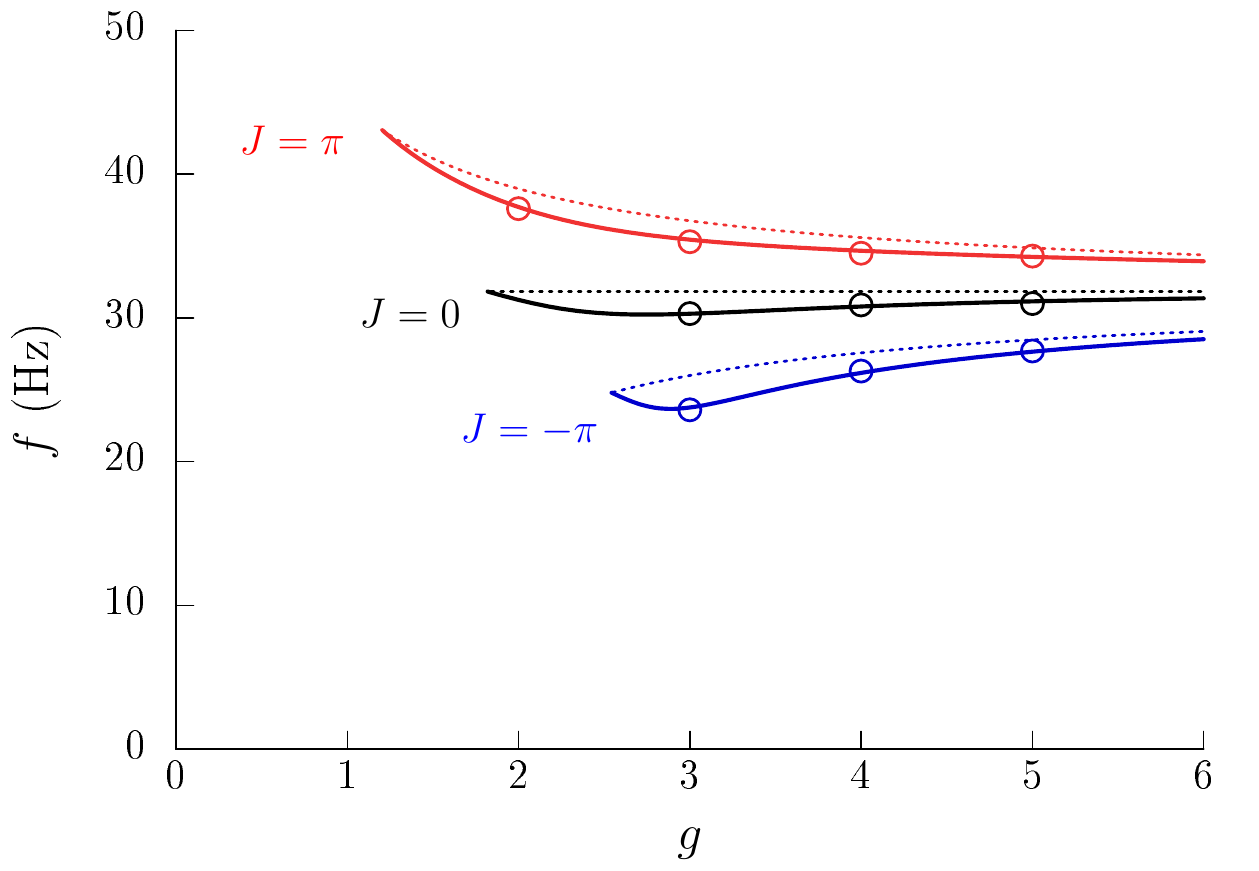}}
\caption{
In the absence of chemical coupling (black), electrical coupling $g$ 
has little effect on the frequency of the oscillations 
Excitatory (red)/Inhibitory (blue) coupling speeds-up/slows-down 
collective oscillations. 
These effects tend to disappear for strong electrical coupling.
The figure shows the frequency of the oscillations $f$ 
as a function of the strength of electrical 
coupling $g$ in networks with Excitation, $J=\pi$, Inhibition, $J=-\pi$, and 
without chemical coupling, $J=0$.
Symbols ($\circ$) are frequencies obtained  
from numerical simulations of a network of $N=10^4$ 
QIF neurons Eqs.~\eqref{qif}. Solid lines are numerically 
obtained  frequencies from the FRE~\eqref{fre}. Dotted lines correspond to 
the Hopf frequency given by 
Eq.~\eqref{fH}. 
Parameters: $\bar \eta=1$, $\tau=10~$ms, and $\Delta=1$.  } 
\label{Fig7}
\end{figure}

\begin{figure}[t]
\centerline{\includegraphics[width=80mm,clip=true]{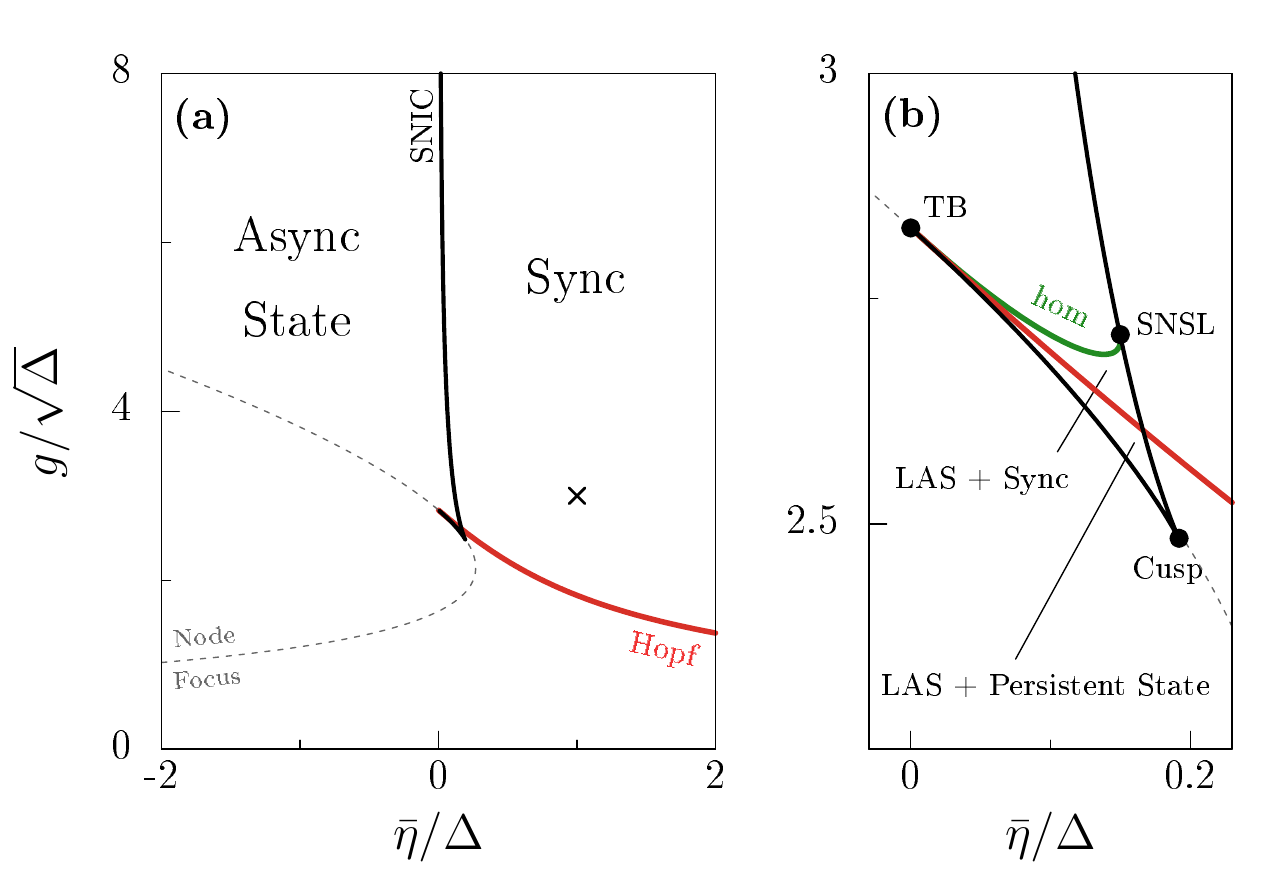}}
\caption{The phase diagram of the FRE~\eqref{fre} for electrical coupling
only ($J=0$) 
is characterized by the presence of three codimension-2 bifurcation points 
(Takens-Bogdanov, TB; Saddle-Node Separatix Loop, SNSL; Cusp), 
all located at $\bar\eta \geq 0$.
The region of synchronization (Sync) is limited by supercritical 
Hopf (red), SNIC (black), and 
homoclinic (green) bifurcations. Dashed line: Focus-Node boundary of the 
Asynchronous State.  
Panel (b) Enlargement of the region near the three codimension-2 points.
There are two small 
regions of bistability between Asynchronous, low-activity-states (LAS) 
and Asynchronous Persistent States, and between LAS and Sync.
Two Saddle-Node (SN) bifurcations are created at a Cusp point, 
at $(1/(3 \sqrt{3}),4\sqrt{2}/3^{3/4})
\approx (0.192,2.482)$. The upper SN line meets
the homoclinic (hom) bifurcation in a SNSL point. 
At this point the upper SN becomes a SNIC 
bifurcation. The other SN bifurcation tangentially meets the homoclinic and the Hopf 
lines at a TB point, at $(0,2\sqrt{2})\approx (0,2.828)$. 
The Hopf boundary 
corresponds to Eq.~\eqref{hopf}. SN/SNIC and Focus/Node 
boundaries are obtained in parametric and explicit form, respectively, in 
\emph{Appendix C}. The homoclinic boundary has been obtained numerically. 
The symbol $\times$
indicates the parameter value considered in Fig.~\ref{Fig4}(a).}
\label{Fig5}
\end{figure}

\begin{figure}[tb]
\centerline{\includegraphics[width=85mm,clip=true]{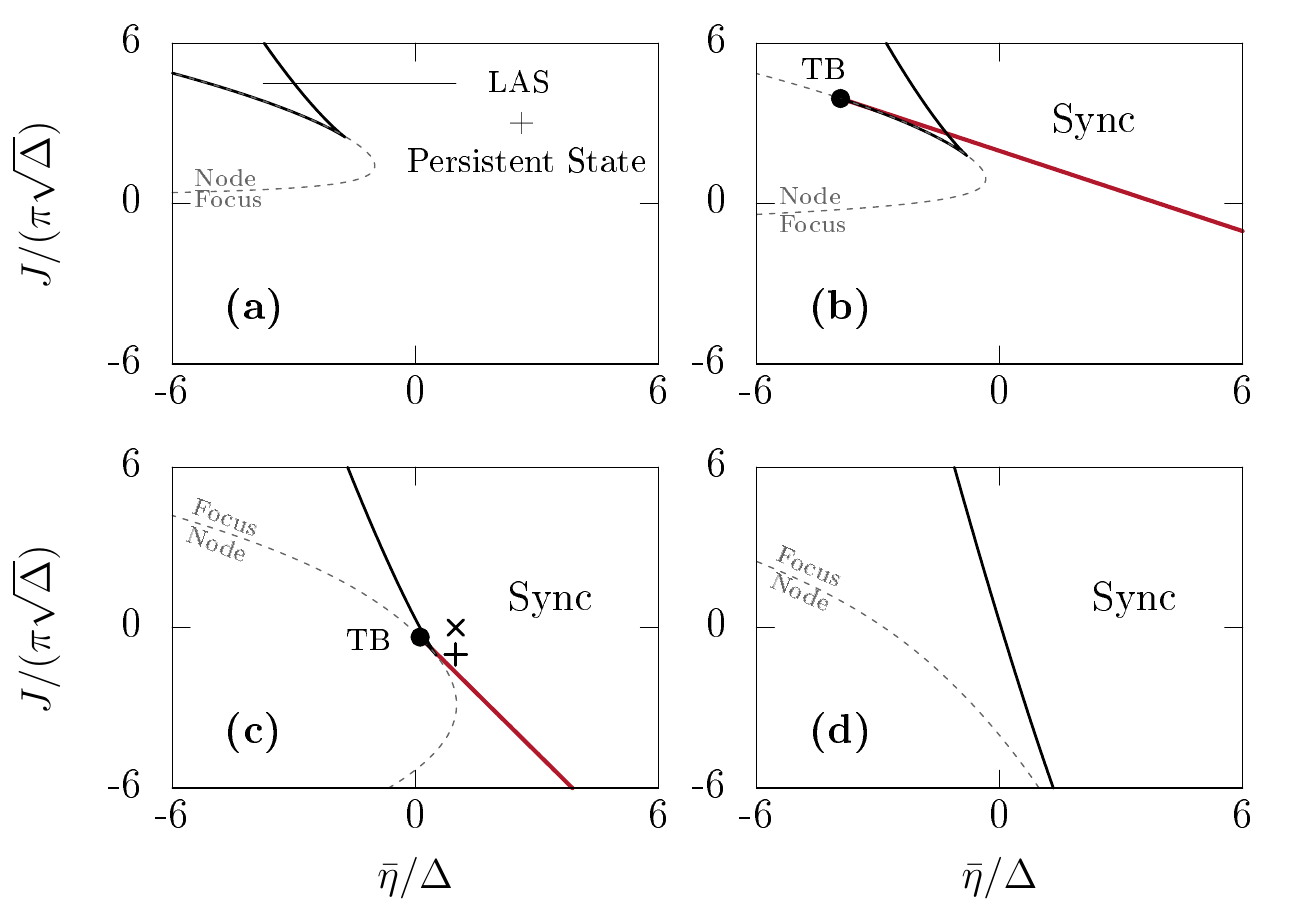}}
\caption{
The phase diagram for networks with only 
chemical coupling, panel (a), is characterized by the presence 
of a Cusp bifurcation point. The inclusion of electrical coupling, panels (b-d),
transforms the Cusp bifurcation scenario into that of Fig.~\ref{Fig5},
characterized by the presence of three codimension-2 bifurcation points.
The panels show the phase diagrams of the FRE~\eqref{fre}
with chemical coupling ($J>0$: Excitatory, $J<0$: Inhibitory)
for (a) $g/\sqrt{\Delta}=0$, (b) $g/\sqrt{\Delta}=1$,
(c) $g/\sqrt{\Delta}=3$, (d) $g/\sqrt{\Delta}=5$.
The Hopf boundaries (red lines) are straight lines given by Eq.~\eqref{hopf}.
SN/SNIC (black lines) and Focus/Node (dashed) boundaries 
are obtained in parametric form in \emph{Appendix C}.
Hopf and SN boundaries meet at a TB bifurcation point.
To lighten the diagrams, Cusp, SNSL and 
homoclinic bifurcations are not shown.
Symbols $\times$ and $+$ indicate the parameter values considered in Fig.~\ref{Fig4}.}
\label{Fig6}
\end{figure}

To confirm the presence of collective oscillations in the original 
network of electrically coupled QIF neurons with dynamics Eq.~\eqref{qif},
we carried out numerical simulations and compared them with 
those of the FRE~\eqref{fre}. Fig.~\ref{Fig4} shows the 
time series of the firing rate in the full and in the reduced system, which display
a very good agreement. In panel (a) we considered a network with electrical 
coupling only. The frequency of the oscillations, $f\approx 30.1$~Hz,
is close to the theoretical value at criticality,
given by Eq.~\eqref{fH}:  
$f_{\text{\scriptsize{H}}}=100/\pi\approx 31.8$~Hz.
Therefore, in absence of chemical coupling and near 
the Hopf bifurcation, the frequency of the oscillations is almost
independent of the coupling strength $g$ and 
closely follows Eq.~\eqref{fH}.
To further test the validity of Eq.~\eqref{fH} far from criticality,   
we numerically evaluated 
the frequency of the limit cycle of the FRE~\eqref{fre} 
(black solid line, Fig.~\ref{Fig7}) 
as the the coupling strength $g$ is increased from the Hopf bifurcation (at $g_H\approx 1.8$). The black dotted line corresponds to 
the Hopf frequency Eq.~\eqref{fH}. We find that the frequency of the limit cycle 
remains close to this for a broad range of $g$ values. 

Hopf instability in networks of electrically coupled QIF neurons 
occurs like the transition to synchronization in the 
Kuramoto model of coupled phase oscillators~\cite{Kur75}. 
Considering $J=0$, we find the main features 
of the Kuramoto transition to collective synchronization:  
($i$) In the limit of weak electrical coupling $g$, the Hopf boundary 
Eq.~\eqref{hopf} can be written as 
\begin{equation}
g_H \approx \frac{2\Delta}{\sqrt{\bar\eta}}.
\label{gcr}
\end{equation}
For $\bar\eta=1$, Eq.~\eqref{gcr} coincides with Kuramoto's critical coupling for synchrony. 
($ii$) As previously discussed, macroscopic oscillations emerge with a 
frequency determined by the most likely value of the 
natural frequencies in the network, see Eq.~\eqref{freq}. For 
the case of the Lorentzian distribution of currents, Eq.~\eqref{lorentzian2},
the most likely value of the frequency is
\begin{equation}
\bar f=\frac{\sqrt{\bar \eta}}{\pi \tau}.
\label{meanf}
\end{equation}
($iii$) The Hopf bifurcation is always supercritical; cf.~\textit{Appendix D}.
Taken together, for $\bar\eta>0$ and given a certain level of heterogeneity $\Delta$,
synchronization occurs ---at a critical coupling approximately given by Eq.~\eqref{gcr}---
with the nucleation of a small cluster of oscillators with 
natural frequencies Eq.~\eqref{freq} near $\bar f$. As  
electrical coupling $g$ is further increased,
more and more oscillators become entrained to the frequency $\bar f$, 
resulting in a continuous and monotonous increase in the amplitude of the oscillations.  
This transition is in contrast to that of networks with inhibitory coupling
and synaptic kinetics and/or delays, where synchrony is only achieved 
for weak heterogeneity and weak coupling, see, e.g.,~\cite{DRM17,DMP18}.

%

The phase diagram depicted in Fig.~\ref{Fig5} characterizes 
the dynamics of the firing rate model Eq.~\eqref{fre} with only electrical coupling. 
The red curve corresponds to the Hopf bifurcation line given by Eq.~\eqref{hopf}. 
According to Eq.~\eqref{fH}, the frequency of the collective 
oscillations approaches zero as $\bar \eta \to 0$. This indicates that the 
Hopf line ends in a Takens-Bogdanov (TB) bifurcation at $\bar \eta=0$, 
see Fig.~\ref{Fig5}(b).
At this codimension-2 point, the Hopf boundary tangentially meets a SN bifurcation and
a homoclinic bifurcation. The homoclinic line 
moves parallel to the Hopf line for a while, it makes a sharp backward turn and then  
tangentially joins onto the upper branch of the SN bifurcation curve 
(two branches of SN bifurcations are created at the Cusp point),
at a saddle-node-separatrix-loop (SNSL) point. 
At this point the SN boundary becomes a SNIC boundary that, together with the Hopf and 
homoclinic lines, encloses the region of synchronization (Sync) featuring
collective oscillations. 
Note that in Fig.~\ref{Fig5}(b) we encounter a very small region of 
bistability between a Low-Activity State (LAS, node) and a persistent 
state (focus). Electrical coupling destabilizes the persistent state almost 
immediately after the SN line, 
leading to another small region of bistability between LAS and a 
small amplitude limit cycle (Sync) ---which disappears in the homoclinic bifurcation.  

Finally, the SNIC curve  
asymptotically approaches  $\bar \eta=0$ as $g/\sqrt{\Delta }\to \infty$
(as suggested by the $N=2$ analysis in Section~\ref{secII_twoneurons}). 
In this 
limit, all neurons are strongly coupled ($g \to \infty$) 
and/or are nearly identical ($\Delta \to 0$) so that they 
behave as a single QIF neuron with input current $\bar \eta$~\footnote{ 
Exactly the same scenario 
is found in systems of globally, sine coupled `active rotators'. 
Such systems are, in fact, closely related to the dynamics~\eqref{qif} 
with $J=0$~\cite{SSK88,ZNF+03,CS08,LCT10}.}.

%
\subsection{Networks with both chemical and electrical coupling}

We finally analyze the dynamics of 
a population of QIF neurons with mixed, chemical and electrical synapses. 
Fig.~\ref{Fig6}(a) presents the possible dynamical regimes of
a population with chemical synapses only, $g=0$.
In contrast to networks with pure electrical coupling, 
where the bifurcation scenario is determined by the presence of three codimension-2 
points, cf.~Fig.~\ref{Fig5}, here there is only a Cusp point, see also~\cite{MPR15}. 
This entails the presence of a persistent state (focus) coexisting 
with an asynchronous, LAS (node) within the cusp-shaped 
region in the top-left corner of Fig.~\ref{Fig6}(a). 
Additionally, the dashed line indicates that the asynchronous state is of focus type in a vast 
region of parameters for excitatory coupling, and always for inhibitory coupling.

Including electrical coupling, $g>0$, yields the Hopf bifurcation given by Eq.~\eqref{hopf}, 
which joins onto the lower branch of
the SN bifurcation curve at a TB point, see Figs.~\ref{Fig6}(b,c). Hence, 
the bifurcation scenario for networks with electrical and chemical synapses 
matches that for networks with electrical synapses only: 
Similar to Fig.~\ref{Fig5}, 
the Hopf line cuts through the cusp-shaped region ---the TB 
bifurcation demarcates the point where the Hopf boundary and the lower SN line intersect.
Then, due to the presence of electrical coupling,
the persistent state becomes only stable 
in a small parameter region confined between the Hopf and the lower SN line, 
see Fig.~\ref{Fig6}(b). As electrical coupling is increased, the TB point approaches the Cusp 
bifurcation, which results in an even smaller range of parameters 
for which the persistent state is stable.
This agrees with numerical results using large networks of noisy, 
conductance-based and QIF neurons, and has been hypothesized to be a possible reason 
why electrical synapses are rarely found 
between excitatory neurons~\cite{Erm06}. 

Returning to the analysis of the FRE~\eqref{fre}, we find for low values of $g$ that synchronization
emerges predominantly for excitatory coupling, $J > 0$, see Fig.~\ref{Fig6}(b). 
As electrical coupling is increased, the Sync region extends to the inhibitory 
region, $J<0$, and to larger values of $\bar \eta$ ---note that in this coupling regime
the emergence of collective oscillations mainly occurs via a 
SNIC bifurcation for excitation and via a Hopf bifurcation for inhibition, see Fig.~\ref{Fig6}(c). 
For even larger electrical coupling, the TB point moves further into the inhibitory region.
That is, 
for strong electrical coupling the $J$-coordinate of the TB bifurcation
rapidly decreases towards minus infinity whereas the other coordinate stays relatively close
to the $\bar \eta=0$ axis
\footnote{
It can be shown that the TB point behaves proportional
to $(\tilde\eta,\tilde{J})_{\text{\scriptsize{TB}}} \propto (g^2,-g^3)$ 
in the limit $g\to\infty$.
}.
The SNIC bifurcation tilts 
towards a vertical line close to the
$\bar \eta=0$ axis, because
strong electrical coupling coerces all neurons to behave as a single QIF neuron 
with common input $\eta=\bar\eta$. Then, 
the SNIC bifurcation becomes the only transition 
between the two possible dynamical regimes, asynchrony or synchrony, see Fig.~\ref{Fig6}(d).

Fig.~\ref{Fig4} shows how the addition of inhibitory coupling 
into a network with only electrical synapses degrades synchrony
---parameters used in Fig.~\ref{Fig4} correspond 
to the symbols shown in Fig.~\ref{Fig6}(c).   
The presence of inhibition clearly slows down the oscillations, 
as predicted by Eq.~\eqref{fH}.

Although Eq.~\eqref{fH} is 
strictly valid only at the Hopf bifurcation, it is a good estimate of the 
frequency of the oscillations of the FRE \eqref{fre}.
Fig.~\ref{Fig7} depicts
the comparison between Eq.~\eqref{fH} as a function of $g$ (dotted lines), 
with the actual frequencies numerically obtained using the FRE \eqref{fre} (solid lines) 
and the QIF network Eq.~\eqref{qif} (symbols). 
In excitatory networks, the oscillations already emerge for weak values of $g$.
In contrast, synchronizing inhibitory networks requires a much larger value of $g$,
i.e. inhibition does not promote synchronization.  
Remarkably, only the presence of chemical coupling allows the frequency of the oscillations 
to deviate from $\bar f$, 
see Eqs.~(\ref{fH},\ref{meanf}). 
Oscillations emerge with $f>\bar f$  for excitation and
with $f<\bar f$ for inhibition, while they remain $f\approx \bar f$ for networks 
with only electrical coupling. 
As $g$ increases, the effects of chemical coupling are gradually washed out 
since an increasing number of neurons are entrained by electrical coupling 
to the most-likely frequency of the uncoupled network, $\bar f$.  
This dependence is well described by Eq.~\eqref{fH}. 
Finally, since 
the level of heterogeneity $\Delta$ degrades synchrony, 
in Eq.~\eqref{fH} this term favors 
the deviation of the frequency from $\bar f$, 
and compensates for the homogenizing effect of electrical coupling. 
In the limit of identical neurons $\Delta \to 0$, the effects 
of instantaneous chemical coupling on the frequency vanish, since 
neurons synchronize in-phase and, 
at the instant of firing, all neurons become refractory.

\section{Conclusion and discussion}
\label{secV}
Firing rate models are  very useful tools for  investigating the dynamics 
of large networks of spiking neurons that interact via  
excitation and inhibition, see e.g.~\cite{WC72,DA01,ET10,Hop84,MBT08,BLS95,HS98,TSO+00,RSR15,MR13,TPM98,RBH05,RM11}.
Remarkably, using a recently proposed approach to derive exact firing rate equations 
for networks of excitatory and/or inhibitory QIF neurons~\cite{MPR15,LBS13}, 
Laing has found that electrical synapses can also be incorporated in the framework of 
firing rate models~\cite{Lai15}. 
Yet, the FRE in~\cite{Lai15} are not exact, 
and their mathematical form makes the analysis intractable.

Here we showed that the FRE
corresponding to a network of QIF neurons with both chemical and electrical 
synapses can be exactly obtained, without the need for any approximation. 
Much in the spirit of firing rate models, the resulting FRE~\eqref{fre} 
are simple in form and highly amenable to analysis. Moreover, 
in {\it Appendix B}, we demonstrate that 
relaxing the approximation invoked in~\cite{Lai15} the 
FRE derived by Laing simplify to our Eqs.~\eqref{fre}.

At first glance, the mathematical form of Eqs.~\eqref{fre} 
already unveils two interesting features of electrical and chemical 
coupling, see also Eq.~\eqref{la0}: 
($i$) Chemical coupling tends to shift 
the center of the distribution of membrane potentials (given by $v$), 
while electrical coupling tends to 
reduce the width of the distribution (given by $r$), potentially 
promoting the emergence of synchronization;
($ii$) While in the original network of 
QIF neurons electrical coupling is mediated by membrane potential differences, at 
the mean-field level the electrical interaction is solely mediated by 
the mean firing rate $r$.

The mathematical analysis of the FRE~\eqref{fre} 
unravels how chemical and electrical coupling shape the dynamics of globally 
coupled populations of QIF neurons with Lorentzian heterogeneity.  Some of our results
were already reported in previous 
work, and confirm the value and the validity of the FRE~\eqref{fre}.
An important conclusion of our study is that 
the presence of electrical coupling, $g \neq 0$, generally 
implies the appearance of a supercritical Hopf bifurcation, see Eq.~\eqref{hopf}.
This Hopf bifurcation meets a SN bifurcating 
line in a codimension-2 TB point, causing a drastic reduction of the region 
of bistability between low-activity and persistent asynchronous states, 
see Figs.~\ref{Fig5} and \ref{Fig6}. 
The Hopf bifurcation destabilizes the persistent state producing 
synchronous oscillations, which then are abolished via a homoclinic bifurcation.
Previous studies of networks of excitatory and inhibitory neurons 
showed that synchrony often destroys persistent states~\cite{HM01,HM03,GLC+01,LC01,KS03}. 
Moreover, the generality of the bifurcation scenario of 
Figs.~\ref{Fig5} and \ref{Fig6} ---characterized by three codimension-2 points, 
TB, Cusp and SNSL---, is confirmed in previous studies analyzing  
closely related systems~\cite{SSK88,ZNF+03,CS08,LCT10}. 

Networks of spiking neurons with strong excitatory coupling display 
robust persistent states. These states emerge at a Cusp bifurcation, see Fig.~\ref{Fig6}(a).
Of particular relevance to our study is the work by Ermentrout~\cite{Erm06}. He
found that electrical coupling tends to synchronize neurons, and that this anihilates  
persistent states via the bifurcation scenario described in Figs.~\ref{Fig5} and ~\ref{Fig6}. 
Persistent activity may underlie important cognitive functions 
such as working memory, and has been suggested as a possible reason for the lack 
of electrical coupling between excitatory neurons~\cite{Erm06}.
According to~\cite{Erm06}, \emph{`the main role for gap junctions 
is to encourage synchronization during rhythmic behavior. 
Synchrony, because it leads to a shared 
refractory period between neurons can lead to the 
extinction of persistent activity'}. 

The Hopf bifurcation is always 
supercritical. In~\cite{OBH09}, Ostojic et al. analyzed the  super- or sub-critical character of the 
Hopf bifurcation in networks of electrically coupled leaky integrate and fire (LIF)
neurons. At variance with QIF neurons, LIF neurons do not have spikes and, hence,
modeling 
electrical coupling requires an additional parameter~\cite{LR03}. 
This parameter enables one 
to adjust the shape of the spikelet elicited in the postsynaptic cell 
due to an action potential in the presynaptic cell. Ostojic et al.~\cite{OBH09}
 found that the Hopf bifurcation is supercritical when the spikelets are 
effectively excitatory, while inhibitory spikelets 
lead to subcritical Hopf bifurcations. For the QIF model, the 
spikelet elicited in a postsynaptic cell by the transmission of a presynaptic spike 
has a net excitatory effect ---see Fig.~\ref{Fig1}(b)---, and hence our result that 
the Hopf bifurcation is supercritical is in agreement with the results in~\cite{OBH09}. 
Yet, we note that our result also includes networks with chemical synapses, and not 
only networks with electrical synapses, as in~\cite{OBH09}. 

Another important result by Ostojic et al.~\cite{OBH09}
is that electrical coupling can 
lead to oscillations even in the presence of strong heterogeneity.
Our Eq.~\eqref{hopf} is consistent with this. 
Kopell and Ermentrout~\cite{KE04} also investigated the robustness of synchrony against 
current heterogeneities in networks with both electrical and inhibitory synapses. 
They found that a small amount of electrical coupling, added to an 
already significant inhibitory coupling, 
can increase synchronization more than a very large increase 
in the inhibitory coupling. In Fig.~\ref{Fig4}
we show that increasing inhibition reduces the amplitude of the oscillations
in a network with $g\neq 0$. 
In addition, Fig.~\ref{Fig6} shows that, for a given value of $g$,
increasing inhibition leads to asynchrony. This level of inhibition increases  
with electrical coupling, in line with the results in~\cite{KE04}.
Two studies~\cite{OBH09,VMG16} 
also investigated the frequency of the emerging oscillations in networks with 
electrical synapses. This frequency 
remains tied to the mean firing rate $f_i$ in the network (i.e. near $\bar f$),
as our Eq.~\eqref{fH} suggests. In Fig.~\ref{Fig7} we confirm that, in 
networks with only electrical synapses, the frequency of the oscillations remains near the most likely $f_i$-value: $\bar f$. 

The result that the Hopf bifurcation is always supercritical, and that the frequency of 
the emerging oscillations is given by $\bar f$ 
evoke the paradigmatic synchronization transition 
in the Kuramoto model~\cite{Kur75}. For weakly 
electrically-coupled networks, we find that the onset of oscillations occurs at 
the Kuramoto's critical coupling for synchrony, Eq.~\eqref{gcr}. 
When considering chemical coupling, 
the frequency of the oscillations deviates from $\bar f$,
increasing/decreasing for excitatory/inhibitory coupling. 
The intensity of this deviation depends on the ratio of chemical to electrical coupling,
as Eq.~\eqref{fH} suggests. Fig.~\ref{Fig7} confirms that strong electrical coupling 
overcomes the effect of excitation/inhibition onto the frequency of the oscillations, 
approximately as dictated by Eq.~\eqref{fH}.

Together with the firing rate model derived in~\cite{Lai15}, 
the FRE~\eqref{fre} constitute a unique example of a firing rate model with both 
electrical and chemical coupling.
The numerical simulations of the original QIF 
network Eq.~\eqref{qif} are in agreement with the FRE~\eqref{fre} 
---see Figs.~\ref{Fig4}, \ref{Fig7}---,
underlining the validity of the reduction method applied.
Interestingly, the fixed points of the FRE~\eqref{fre} 
with chemical synapses ($g=0$,~$J\neq 0$) can be cast in the form 
of a traditional firing rate model     
\begin{equation}
r_*=\Phi(\bar \eta+ J \tau r_*),
\label{WC}
\end{equation}
where $\Phi(x)=\sqrt{x+\sqrt{x^2+\Delta^2}}/(\sqrt{2}\pi\tau)$ is the so-called 
transfer function of the heterogeneous QIF network 
~\cite{ERA+17,DRM17,DMP18}.
The FRE~\eqref{fre} with electrical synapses ($g\neq 0$), however,
cannot be written in the form of Eq.~\eqref{WC}. Therefore, the link
pointed by Eq.~\eqref{WC}
between traditional firing rate models and Eqs.~\eqref{fre} is lost when  
electrical coupling is considered.

\section*{Acknowledgements}
This work was supported by ITN COSMOS funded by the EU Horizon 2020 
Research and Innovation programme under the 
Marie Sk\l{}odowska-Curie Grant Agreement No. 642563.
EM acknowledges support by the Spanish Ministry of
Economy and Competitiveness under Project No. PSI2016-75688-P.


\begin{thebibliography}{78}
\expandafter\ifx\csname natexlab\endcsname\relax\def\natexlab#1{#1}\fi
\expandafter\ifx\csname bibnamefont\endcsname\relax
  \def\bibnamefont#1{#1}\fi
\expandafter\ifx\csname bibfnamefont\endcsname\relax
  \def\bibfnamefont#1{#1}\fi
\expandafter\ifx\csname citenamefont\endcsname\relax
  \def\citenamefont#1{#1}\fi
\expandafter\ifx\csname url\endcsname\relax
  \def\url#1{\texttt{#1}}\fi
\expandafter\ifx\csname urlprefix\endcsname\relax\def\urlprefix{URL }\fi
\providecommand{\bibinfo}[2]{#2}
\providecommand{\eprint}[2][]{\url{#2}}

\bibitem[{\citenamefont{Wang}(2010)}]{Wan10}
\bibinfo{author}{\bibfnamefont{X.-J.} \bibnamefont{Wang}},
  \bibinfo{journal}{Physiological Reviews} \textbf{\bibinfo{volume}{90}},
  \bibinfo{pages}{1195} (\bibinfo{year}{2010}).

\bibitem[{\citenamefont{Nagy et~al.}(2018)\citenamefont{Nagy, Pereda, and
  Rash}}]{NPR18}
\bibinfo{author}{\bibfnamefont{J.~I.} \bibnamefont{Nagy}},
  \bibinfo{author}{\bibfnamefont{A.~E.} \bibnamefont{Pereda}},
  \bibnamefont{and} \bibinfo{author}{\bibfnamefont{J.~E.} \bibnamefont{Rash}},
  \bibinfo{journal}{Biochimica et Biophysica Acta (BBA) - Biomembranes}
  \textbf{\bibinfo{volume}{1860}}, \bibinfo{pages}{102 }
  (\bibinfo{year}{2018}).

\bibitem[{\citenamefont{Connors}(2017)}]{Con17}
\bibinfo{author}{\bibfnamefont{B.~W.} \bibnamefont{Connors}},
  \bibinfo{journal}{Developmental Neurobiology} \textbf{\bibinfo{volume}{77}},
  \bibinfo{pages}{610} (\bibinfo{year}{2017}).

\bibitem[{\citenamefont{Traub et~al.}(2018)\citenamefont{Traub, Whittington,
  Guti{\'e}rrez, and Draguhn}}]{TWG+18}
\bibinfo{author}{\bibfnamefont{R.~D.} \bibnamefont{Traub}},
  \bibinfo{author}{\bibfnamefont{M.~A.} \bibnamefont{Whittington}},
  \bibinfo{author}{\bibfnamefont{R.}~\bibnamefont{Guti{\'e}rrez}},
  \bibnamefont{and} \bibinfo{author}{\bibfnamefont{A.}~\bibnamefont{Draguhn}},
  \bibinfo{journal}{Cell and Tissue Research} \textbf{\bibinfo{volume}{373}},
  \bibinfo{pages}{671} (\bibinfo{year}{2018}).

\bibitem[{\citenamefont{Bennett and Zukin}(2004)}]{BZ04}
\bibinfo{author}{\bibfnamefont{M.~V.} \bibnamefont{Bennett}} \bibnamefont{and}
  \bibinfo{author}{\bibfnamefont{R.}~\bibnamefont{Zukin}},
  \bibinfo{journal}{Neuron} \textbf{\bibinfo{volume}{41}}, \bibinfo{pages}{495
  } (\bibinfo{year}{2004}).

\bibitem[{\citenamefont{Connors and Long}(2004)}]{CL04}
\bibinfo{author}{\bibfnamefont{B.~W.} \bibnamefont{Connors}} \bibnamefont{and}
  \bibinfo{author}{\bibfnamefont{M.~A.} \bibnamefont{Long}},
  \bibinfo{journal}{Annual Review of Neuroscience}
  \textbf{\bibinfo{volume}{27}}, \bibinfo{pages}{393} (\bibinfo{year}{2004}).

\bibitem[{\citenamefont{Whittington et~al.}(2000)\citenamefont{Whittington,
  Traub, Kopell, Ermentrout, and Buhl}}]{WTK+00}
\bibinfo{author}{\bibfnamefont{M.}~\bibnamefont{Whittington}},
  \bibinfo{author}{\bibfnamefont{R.}~\bibnamefont{Traub}},
  \bibinfo{author}{\bibfnamefont{N.}~\bibnamefont{Kopell}},
  \bibinfo{author}{\bibfnamefont{B.}~\bibnamefont{Ermentrout}},
  \bibnamefont{and} \bibinfo{author}{\bibfnamefont{E.}~\bibnamefont{Buhl}},
  \bibinfo{journal}{Int. Journal of Psychophysiol.}
  \textbf{\bibinfo{volume}{38}}, \bibinfo{pages}{315 } (\bibinfo{year}{2000}),
  ISSN \bibinfo{issn}{0167-8760}.

\bibitem[{\citenamefont{Kopell and Ermentrout}(2004)}]{KE04}
\bibinfo{author}{\bibfnamefont{N.}~\bibnamefont{Kopell}} \bibnamefont{and}
  \bibinfo{author}{\bibfnamefont{B.}~\bibnamefont{Ermentrout}},
  \bibinfo{journal}{Proceedings of the National Academy of Sciences}
  \textbf{\bibinfo{volume}{101}}, \bibinfo{pages}{15482}
  (\bibinfo{year}{2004}).

\bibitem[{\citenamefont{Pfeuty et~al.}(2007)\citenamefont{Pfeuty, Golomb, Mato,
  and Hansel}}]{PGM+07}
\bibinfo{author}{\bibfnamefont{B.}~\bibnamefont{Pfeuty}},
  \bibinfo{author}{\bibfnamefont{D.}~\bibnamefont{Golomb}},
  \bibinfo{author}{\bibfnamefont{G.}~\bibnamefont{Mato}}, \bibnamefont{and}
  \bibinfo{author}{\bibfnamefont{D.}~\bibnamefont{Hansel}},
  \bibinfo{journal}{Frontiers in Computational Neuroscience}
  \textbf{\bibinfo{volume}{1}}, \bibinfo{pages}{8} (\bibinfo{year}{2007}), ISSN
  \bibinfo{issn}{1662-5188}.

\bibitem[{\citenamefont{Ermentrout}(2006)}]{Erm06}
\bibinfo{author}{\bibfnamefont{B.}~\bibnamefont{Ermentrout}},
  \bibinfo{journal}{Phys. Rev. E} \textbf{\bibinfo{volume}{74}},
  \bibinfo{pages}{031918} (\bibinfo{year}{2006}).

\bibitem[{\citenamefont{Viriyopase et~al.}(2016)\citenamefont{Viriyopase,
  Memmesheimer, and Gielen}}]{VMG16}
\bibinfo{author}{\bibfnamefont{A.}~\bibnamefont{Viriyopase}},
  \bibinfo{author}{\bibfnamefont{R.-M.} \bibnamefont{Memmesheimer}},
  \bibnamefont{and} \bibinfo{author}{\bibfnamefont{S.}~\bibnamefont{Gielen}},
  \bibinfo{journal}{Journal of Neurophysiology} \textbf{\bibinfo{volume}{116}},
  \bibinfo{pages}{232} (\bibinfo{year}{2016}).

\bibitem[{\citenamefont{Holzbecher and Kempter}(2018)}]{HK18}
\bibinfo{author}{\bibfnamefont{A.}~\bibnamefont{Holzbecher}} \bibnamefont{and}
  \bibinfo{author}{\bibfnamefont{R.}~\bibnamefont{Kempter}},
  \bibinfo{journal}{European Journal of Neuroscience}
  \textbf{\bibinfo{volume}{48}}, \bibinfo{pages}{3446} (\bibinfo{year}{2018}).

\bibitem[{\citenamefont{Guo et~al.}(2012)\citenamefont{Guo, Wang, and
  Perc}}]{GWP12}
\bibinfo{author}{\bibfnamefont{D.}~\bibnamefont{Guo}},
  \bibinfo{author}{\bibfnamefont{Q.}~\bibnamefont{Wang}}, \bibnamefont{and}
  \bibinfo{author}{\bibfnamefont{M.~c.~v.} \bibnamefont{Perc}},
  \bibinfo{journal}{Phys. Rev. E} \textbf{\bibinfo{volume}{85}},
  \bibinfo{pages}{061905} (\bibinfo{year}{2012}).

\bibitem[{\citenamefont{Tchumatchenko and Clopath}(2014)}]{TC14}
\bibinfo{author}{\bibfnamefont{T.}~\bibnamefont{Tchumatchenko}}
  \bibnamefont{and} \bibinfo{author}{\bibfnamefont{C.}~\bibnamefont{Clopath}},
  \bibinfo{journal}{Nature Communications} \textbf{\bibinfo{volume}{5}},
  \bibinfo{pages}{5512} (\bibinfo{year}{2014}).

\bibitem[{\citenamefont{Lewis and Rinzel}(2003)}]{LR03}
\bibinfo{author}{\bibfnamefont{T.~J.} \bibnamefont{Lewis}} \bibnamefont{and}
  \bibinfo{author}{\bibfnamefont{J.}~\bibnamefont{Rinzel}},
  \bibinfo{journal}{Journal of Computational Neuroscience}
  \textbf{\bibinfo{volume}{14}}, \bibinfo{pages}{283} (\bibinfo{year}{2003}).

\bibitem[{\citenamefont{Chow and Kopell}(2000)}]{CK00}
\bibinfo{author}{\bibfnamefont{C.~C.} \bibnamefont{Chow}} \bibnamefont{and}
  \bibinfo{author}{\bibfnamefont{N.}~\bibnamefont{Kopell}},
  \bibinfo{journal}{Neural Computation} \textbf{\bibinfo{volume}{12}},
  \bibinfo{pages}{1643} (\bibinfo{year}{2000}).

\bibitem[{\citenamefont{Ostojic et~al.}(2009)\citenamefont{Ostojic, Brunel, and
  Hakim}}]{OBH09}
\bibinfo{author}{\bibfnamefont{S.}~\bibnamefont{Ostojic}},
  \bibinfo{author}{\bibfnamefont{N.}~\bibnamefont{Brunel}}, \bibnamefont{and}
  \bibinfo{author}{\bibfnamefont{V.}~\bibnamefont{Hakim}},
  \bibinfo{journal}{Journal of Computational Neuroscience}
  \textbf{\bibinfo{volume}{26}}, \bibinfo{pages}{369} (\bibinfo{year}{2009}),
  ISSN \bibinfo{issn}{1573-6873}.

\bibitem[{\citenamefont{Pfeuty et~al.}(2003)\citenamefont{Pfeuty, Mato, Golomb,
  and Hansel}}]{PMG+03}
\bibinfo{author}{\bibfnamefont{B.}~\bibnamefont{Pfeuty}},
  \bibinfo{author}{\bibfnamefont{G.}~\bibnamefont{Mato}},
  \bibinfo{author}{\bibfnamefont{D.}~\bibnamefont{Golomb}}, \bibnamefont{and}
  \bibinfo{author}{\bibfnamefont{D.}~\bibnamefont{Hansel}},
  \bibinfo{journal}{Journal of Neuroscience} \textbf{\bibinfo{volume}{23}},
  \bibinfo{pages}{6280} (\bibinfo{year}{2003}).

\bibitem[{\citenamefont{Coombes}(2008)}]{Coo08}
\bibinfo{author}{\bibfnamefont{S.}~\bibnamefont{Coombes}},
  \bibinfo{journal}{SIAM Journal on Applied Dynamical Systems}
  \textbf{\bibinfo{volume}{7}}, \bibinfo{pages}{1101} (\bibinfo{year}{2008}).

\bibitem[{\citenamefont{Mancilla et~al.}(2007)\citenamefont{Mancilla, Lewis,
  Pinto, Rinzel, and Connors}}]{MLP+07}
\bibinfo{author}{\bibfnamefont{J.~G.} \bibnamefont{Mancilla}},
  \bibinfo{author}{\bibfnamefont{T.~J.} \bibnamefont{Lewis}},
  \bibinfo{author}{\bibfnamefont{D.~J.} \bibnamefont{Pinto}},
  \bibinfo{author}{\bibfnamefont{J.}~\bibnamefont{Rinzel}}, \bibnamefont{and}
  \bibinfo{author}{\bibfnamefont{B.~W.} \bibnamefont{Connors}},
  \bibinfo{journal}{Journal of Neuroscience} \textbf{\bibinfo{volume}{27}},
  \bibinfo{pages}{2058} (\bibinfo{year}{2007}).

\bibitem[{\citenamefont{Wilson and Cowan}(1972)}]{WC72}
\bibinfo{author}{\bibfnamefont{H.~R.} \bibnamefont{Wilson}} \bibnamefont{and}
  \bibinfo{author}{\bibfnamefont{J.~D.} \bibnamefont{Cowan}},
  \bibinfo{journal}{Biophys. J.} \textbf{\bibinfo{volume}{12}},
  \bibinfo{pages}{1} (\bibinfo{year}{1972}).

\bibitem[{\citenamefont{Dayan and Abbott}(2001)}]{DA01}
\bibinfo{author}{\bibfnamefont{P.}~\bibnamefont{Dayan}} \bibnamefont{and}
  \bibinfo{author}{\bibfnamefont{L.~F.} \bibnamefont{Abbott}},
  \emph{\bibinfo{title}{Theoretical neuroscience}}
  (\bibinfo{publisher}{Cambridge, MA: MIT Press}, \bibinfo{year}{2001}).

\bibitem[{\citenamefont{Ermentrout and Terman}(2010)}]{ET10}
\bibinfo{author}{\bibfnamefont{G.~B.} \bibnamefont{Ermentrout}}
  \bibnamefont{and} \bibinfo{author}{\bibfnamefont{D.~H.}
  \bibnamefont{Terman}}, \emph{\bibinfo{title}{Mathematical foundations of
  neuroscience}}, vol.~\bibinfo{volume}{64} (\bibinfo{publisher}{Springer},
  \bibinfo{year}{2010}).

\bibitem[{\citenamefont{Hopfield}(1984)}]{Hop84}
\bibinfo{author}{\bibfnamefont{J.~J.} \bibnamefont{Hopfield}},
  \bibinfo{journal}{Proceedings of the national academy of sciences}
  \textbf{\bibinfo{volume}{81}}, \bibinfo{pages}{3088} (\bibinfo{year}{1984}).

\bibitem[{\citenamefont{Mongillo et~al.}(2008)\citenamefont{Mongillo, Barak,
  and Tsodyks}}]{MBT08}
\bibinfo{author}{\bibfnamefont{G.}~\bibnamefont{Mongillo}},
  \bibinfo{author}{\bibfnamefont{O.}~\bibnamefont{Barak}}, \bibnamefont{and}
  \bibinfo{author}{\bibfnamefont{M.}~\bibnamefont{Tsodyks}},
  \bibinfo{journal}{Science} \textbf{\bibinfo{volume}{319}},
  \bibinfo{pages}{1543} (\bibinfo{year}{2008}).

\bibitem[{\citenamefont{Ben-Yishai et~al.}(1995)\citenamefont{Ben-Yishai,
  Bar-Or, and Sompolinsky}}]{BLS95}
\bibinfo{author}{\bibfnamefont{R.}~\bibnamefont{Ben-Yishai}},
  \bibinfo{author}{\bibfnamefont{R.~L.} \bibnamefont{Bar-Or}},
  \bibnamefont{and}
  \bibinfo{author}{\bibfnamefont{H.}~\bibnamefont{Sompolinsky}},
  \bibinfo{journal}{Proc. Nat. Acad. Sci.} \textbf{\bibinfo{volume}{92}},
  \bibinfo{pages}{3844} (\bibinfo{year}{1995}).

\bibitem[{\citenamefont{Hansel and Sompolinsky}(1998)}]{HS98}
\bibinfo{author}{\bibfnamefont{D.}~\bibnamefont{Hansel}} \bibnamefont{and}
  \bibinfo{author}{\bibfnamefont{H.}~\bibnamefont{Sompolinsky}}, in
  \emph{\bibinfo{booktitle}{Methods in Neuronal Modelling: From Ions to
  Networks}}, edited by \bibinfo{editor}{\bibfnamefont{C.}~\bibnamefont{Koch}}
  \bibnamefont{and} \bibinfo{editor}{\bibfnamefont{I.}~\bibnamefont{Segev}}
  (\bibinfo{publisher}{MIT Press}, \bibinfo{address}{Cambridge},
  \bibinfo{year}{1998}), pp. \bibinfo{pages}{499--567}.

\bibitem[{\citenamefont{Tabak et~al.}(2000)\citenamefont{Tabak, Senn,
  O’Donovan, and Rinzel}}]{TSO+00}
\bibinfo{author}{\bibfnamefont{J.}~\bibnamefont{Tabak}},
  \bibinfo{author}{\bibfnamefont{W.}~\bibnamefont{Senn}},
  \bibinfo{author}{\bibfnamefont{M.~J.} \bibnamefont{O’Donovan}},
  \bibnamefont{and} \bibinfo{author}{\bibfnamefont{J.}~\bibnamefont{Rinzel}},
  \bibinfo{journal}{J. Neurosci.} \textbf{\bibinfo{volume}{20}},
  \bibinfo{pages}{3041} (\bibinfo{year}{2000}).

\bibitem[{\citenamefont{Rankin et~al.}(2015)\citenamefont{Rankin, Sussman, and
  Rinzel}}]{RSR15}
\bibinfo{author}{\bibfnamefont{J.}~\bibnamefont{Rankin}},
  \bibinfo{author}{\bibfnamefont{E.}~\bibnamefont{Sussman}}, \bibnamefont{and}
  \bibinfo{author}{\bibfnamefont{J.}~\bibnamefont{Rinzel}},
  \bibinfo{journal}{PLoS Computational Biology} \textbf{\bibinfo{volume}{11}},
  \bibinfo{pages}{1} (\bibinfo{year}{2015}).

\bibitem[{\citenamefont{Mart{\'\i} and Rinzel}(2013)}]{MR13}
\bibinfo{author}{\bibfnamefont{D.}~\bibnamefont{Mart{\'\i}}} \bibnamefont{and}
  \bibinfo{author}{\bibfnamefont{J.}~\bibnamefont{Rinzel}},
  \bibinfo{journal}{Neural computation} \textbf{\bibinfo{volume}{25}},
  \bibinfo{pages}{1} (\bibinfo{year}{2013}).

\bibitem[{\citenamefont{Tsodyks~M.}(1998)}]{TPM98}
\bibinfo{author}{\bibfnamefont{M.~H.} \bibnamefont{Tsodyks~M.},
  \bibfnamefont{Pawelzik~K.}}, \bibinfo{journal}{Neural Comput.}
  \textbf{\bibinfo{volume}{10}}, \bibinfo{pages}{821} (\bibinfo{year}{1998}).

\bibitem[{\citenamefont{Roxin et~al.}(2005)\citenamefont{Roxin, Brunel, and
  Hansel}}]{RBH05}
\bibinfo{author}{\bibfnamefont{A.}~\bibnamefont{Roxin}},
  \bibinfo{author}{\bibfnamefont{N.}~\bibnamefont{Brunel}}, \bibnamefont{and}
  \bibinfo{author}{\bibfnamefont{D.}~\bibnamefont{Hansel}},
  \bibinfo{journal}{Phys. Rev. Lett.} \textbf{\bibinfo{volume}{94}},
  \bibinfo{pages}{238103} (\bibinfo{year}{2005}).

\bibitem[{\citenamefont{Roxin and Montbri{\'o}}(2011)}]{RM11}
\bibinfo{author}{\bibfnamefont{A.}~\bibnamefont{Roxin}} \bibnamefont{and}
  \bibinfo{author}{\bibfnamefont{E.}~\bibnamefont{Montbri{\'o}}},
  \bibinfo{journal}{Physica D} \textbf{\bibinfo{volume}{240}},
  \bibinfo{pages}{323} (\bibinfo{year}{2011}).

\bibitem[{\citenamefont{Montbri\'o et~al.}(2015)\citenamefont{Montbri\'o,
  Paz\'o, and Roxin}}]{MPR15}
\bibinfo{author}{\bibfnamefont{E.}~\bibnamefont{Montbri\'o}},
  \bibinfo{author}{\bibfnamefont{D.}~\bibnamefont{Paz\'o}}, \bibnamefont{and}
  \bibinfo{author}{\bibfnamefont{A.}~\bibnamefont{Roxin}},
  \bibinfo{journal}{Phys. Rev. X} \textbf{\bibinfo{volume}{5}},
  \bibinfo{pages}{021028} (\bibinfo{year}{2015}).

\bibitem[{\citenamefont{Ott and Antonsen}(2008)}]{OA08}
\bibinfo{author}{\bibfnamefont{E.}~\bibnamefont{Ott}} \bibnamefont{and}
  \bibinfo{author}{\bibfnamefont{T.~M.} \bibnamefont{Antonsen}},
  \bibinfo{journal}{Chaos} \textbf{\bibinfo{volume}{18}}, \bibinfo{eid}{037113}
  (\bibinfo{year}{2008}).

\bibitem[{\citenamefont{Ott and Antonsen}(2009)}]{OA09}
\bibinfo{author}{\bibfnamefont{E.}~\bibnamefont{Ott}} \bibnamefont{and}
  \bibinfo{author}{\bibfnamefont{T.~M.} \bibnamefont{Antonsen}},
  \bibinfo{journal}{Chaos} \textbf{\bibinfo{volume}{19}}, \bibinfo{eid}{023117}
  (\bibinfo{year}{2009}).

\bibitem[{\citenamefont{Ott et~al.}(2011)\citenamefont{Ott, Hunt, and
  Antonsen}}]{OHA11}
\bibinfo{author}{\bibfnamefont{E.}~\bibnamefont{Ott}},
  \bibinfo{author}{\bibfnamefont{B.~R.} \bibnamefont{Hunt}}, \bibnamefont{and}
  \bibinfo{author}{\bibfnamefont{T.~M.} \bibnamefont{Antonsen}},
  \bibinfo{journal}{Chaos} \textbf{\bibinfo{volume}{21}}, \bibinfo{eid}{025112}
  (\bibinfo{year}{2011}).

\bibitem[{\citenamefont{Pikovsky and Rosenblum}(2008)}]{PR08}
\bibinfo{author}{\bibfnamefont{A.}~\bibnamefont{Pikovsky}} \bibnamefont{and}
  \bibinfo{author}{\bibfnamefont{M.}~\bibnamefont{Rosenblum}},
  \bibinfo{journal}{Phys. Rev. Lett.} \textbf{\bibinfo{volume}{101}},
  \bibinfo{pages}{264103} (\bibinfo{year}{2008}).

\bibitem[{\citenamefont{Marvel et~al.}(2009)\citenamefont{Marvel, Mirollo, and
  Strogatz}}]{MMS09}
\bibinfo{author}{\bibfnamefont{S.~A.} \bibnamefont{Marvel}},
  \bibinfo{author}{\bibfnamefont{R.~E.} \bibnamefont{Mirollo}},
  \bibnamefont{and} \bibinfo{author}{\bibfnamefont{S.~H.}
  \bibnamefont{Strogatz}}, \bibinfo{journal}{Chaos: An Interdisciplinary
  Journal of Nonlinear Science} \textbf{\bibinfo{volume}{19}},
  \bibinfo{eid}{043104} (\bibinfo{year}{2009}).

\bibitem[{\citenamefont{Pikovsky and Rosenblum}(2011)}]{PR11}
\bibinfo{author}{\bibfnamefont{A.}~\bibnamefont{Pikovsky}} \bibnamefont{and}
  \bibinfo{author}{\bibfnamefont{M.}~\bibnamefont{Rosenblum}},
  \bibinfo{journal}{Physica D} \textbf{\bibinfo{volume}{240}},
  \bibinfo{pages}{872 } (\bibinfo{year}{2011}).

\bibitem[{\citenamefont{Pietras and Daffertshofer}(2016)}]{PD16}
\bibinfo{author}{\bibfnamefont{B.}~\bibnamefont{Pietras}} \bibnamefont{and}
  \bibinfo{author}{\bibfnamefont{A.}~\bibnamefont{Daffertshofer}},
  \bibinfo{journal}{Chaos: An Interdisciplinary Journal of Nonlinear Science}
  \textbf{\bibinfo{volume}{26}}, \bibinfo{pages}{103101}
  (\bibinfo{year}{2016}).

\bibitem[{\citenamefont{Luke et~al.}(2013)\citenamefont{Luke, Barreto, and
  So}}]{LBS13}
\bibinfo{author}{\bibfnamefont{T.~B.} \bibnamefont{Luke}},
  \bibinfo{author}{\bibfnamefont{E.}~\bibnamefont{Barreto}}, \bibnamefont{and}
  \bibinfo{author}{\bibfnamefont{P.}~\bibnamefont{So}},
  \bibinfo{journal}{Neural Comput.} \textbf{\bibinfo{volume}{25}},
  \bibinfo{pages}{3207} (\bibinfo{year}{2013}).

\bibitem[{\citenamefont{So et~al.}(2014)\citenamefont{So, Luke, and
  Barreto}}]{SLB14}
\bibinfo{author}{\bibfnamefont{P.}~\bibnamefont{So}},
  \bibinfo{author}{\bibfnamefont{T.~B.} \bibnamefont{Luke}}, \bibnamefont{and}
  \bibinfo{author}{\bibfnamefont{E.}~\bibnamefont{Barreto}},
  \bibinfo{journal}{Physica D} \textbf{\bibinfo{volume}{267}},
  \bibinfo{pages}{16} (\bibinfo{year}{2014}).

\bibitem[{\citenamefont{Laing}(2014)}]{Lai14}
\bibinfo{author}{\bibfnamefont{C.~R.} \bibnamefont{Laing}},
  \bibinfo{journal}{Phys. Rev. E} \textbf{\bibinfo{volume}{90}},
  \bibinfo{pages}{010901} (\bibinfo{year}{2014}).

\bibitem[{\citenamefont{Paz\'o and Montbri\'o}(2016)}]{PM16}
\bibinfo{author}{\bibfnamefont{D.}~\bibnamefont{Paz\'o}} \bibnamefont{and}
  \bibinfo{author}{\bibfnamefont{E.}~\bibnamefont{Montbri\'o}},
  \bibinfo{journal}{Phys. Rev. Lett.} \textbf{\bibinfo{volume}{116}},
  \bibinfo{pages}{238101} (\bibinfo{year}{2016}).

\bibitem[{\citenamefont{Ratas and Pyragas}(2016)}]{RP16}
\bibinfo{author}{\bibfnamefont{I.}~\bibnamefont{Ratas}} \bibnamefont{and}
  \bibinfo{author}{\bibfnamefont{K.}~\bibnamefont{Pyragas}},
  \bibinfo{journal}{Phys. Rev. E} \textbf{\bibinfo{volume}{94}},
  \bibinfo{pages}{032215} (\bibinfo{year}{2016}).

\bibitem[{\citenamefont{Roulet and Mindlin}(2016)}]{RM16}
\bibinfo{author}{\bibfnamefont{J.}~\bibnamefont{Roulet}} \bibnamefont{and}
  \bibinfo{author}{\bibfnamefont{G.~B.} \bibnamefont{Mindlin}},
  \bibinfo{journal}{Chaos: An Interdisciplinary Journal of Nonlinear Science}
  \textbf{\bibinfo{volume}{26}}, \bibinfo{pages}{093104}
  (\bibinfo{year}{2016}).

\bibitem[{\citenamefont{Devalle et~al.}(2018)\citenamefont{Devalle, Montbri\'o,
  and Paz\'o}}]{DMP18}
\bibinfo{author}{\bibfnamefont{F.}~\bibnamefont{Devalle}},
  \bibinfo{author}{\bibfnamefont{E.}~\bibnamefont{Montbri\'o}},
  \bibnamefont{and} \bibinfo{author}{\bibfnamefont{D.}~\bibnamefont{Paz\'o}},
  \bibinfo{journal}{Phys. Rev. E} \textbf{\bibinfo{volume}{98}},
  \bibinfo{pages}{042214} (\bibinfo{year}{2018}).

\bibitem[{\citenamefont{Devalle et~al.}(2017)\citenamefont{Devalle, Roxin, and
  Montbri\'o}}]{DRM17}
\bibinfo{author}{\bibfnamefont{F.}~\bibnamefont{Devalle}},
  \bibinfo{author}{\bibfnamefont{A.}~\bibnamefont{Roxin}}, \bibnamefont{and}
  \bibinfo{author}{\bibfnamefont{E.}~\bibnamefont{Montbri\'o}},
  \bibinfo{journal}{PLoS Computational Biology} \textbf{\bibinfo{volume}{13}}
  (\bibinfo{year}{2017}).

\bibitem[{\citenamefont{Esnaola-Acebes
  et~al.}(2017)\citenamefont{Esnaola-Acebes, Roxin, Avitabile, and
  Montbri\'o}}]{ERA+17}
\bibinfo{author}{\bibfnamefont{J.~M.} \bibnamefont{Esnaola-Acebes}},
  \bibinfo{author}{\bibfnamefont{A.}~\bibnamefont{Roxin}},
  \bibinfo{author}{\bibfnamefont{D.}~\bibnamefont{Avitabile}},
  \bibnamefont{and}
  \bibinfo{author}{\bibfnamefont{E.}~\bibnamefont{Montbri\'o}},
  \bibinfo{journal}{Phys. Rev. E} \textbf{\bibinfo{volume}{96}},
  \bibinfo{pages}{052407} (\bibinfo{year}{2017}).

\bibitem[{\citenamefont{Ratas and Pyragas}(2017)}]{RP17}
\bibinfo{author}{\bibfnamefont{I.}~\bibnamefont{Ratas}} \bibnamefont{and}
  \bibinfo{author}{\bibfnamefont{K.}~\bibnamefont{Pyragas}},
  \bibinfo{journal}{Phys. Rev. E} \textbf{\bibinfo{volume}{96}},
  \bibinfo{pages}{042212} (\bibinfo{year}{2017}).

\bibitem[{\citenamefont{Dumont et~al.}(2017)\citenamefont{Dumont, Ermentrout,
  and Gutkin}}]{DEG17}
\bibinfo{author}{\bibfnamefont{G.}~\bibnamefont{Dumont}},
  \bibinfo{author}{\bibfnamefont{G.~B.} \bibnamefont{Ermentrout}},
  \bibnamefont{and} \bibinfo{author}{\bibfnamefont{B.}~\bibnamefont{Gutkin}},
  \bibinfo{journal}{Phys. Rev. E} \textbf{\bibinfo{volume}{96}},
  \bibinfo{pages}{042311} (\bibinfo{year}{2017}).

\bibitem[{\citenamefont{Byrne et~al.}(2017)\citenamefont{Byrne, Brookes, and
  Coombes}}]{BBC17}
\bibinfo{author}{\bibfnamefont{{\'A}.}~\bibnamefont{Byrne}},
  \bibinfo{author}{\bibfnamefont{M.~J.} \bibnamefont{Brookes}},
  \bibnamefont{and} \bibinfo{author}{\bibfnamefont{S.}~\bibnamefont{Coombes}},
  \bibinfo{journal}{Journal of Computational Neuroscience}
  \textbf{\bibinfo{volume}{43}}, \bibinfo{pages}{143} (\bibinfo{year}{2017}).

\bibitem[{\citenamefont{Laing}(2018)}]{Lai18}
\bibinfo{author}{\bibfnamefont{C.~R.} \bibnamefont{Laing}},
  \bibinfo{journal}{The Journal of Mathematical Neuroscience}
  \textbf{\bibinfo{volume}{8}}, \bibinfo{pages}{4} (\bibinfo{year}{2018}), ISSN
  \bibinfo{issn}{2190-8567}.

\bibitem[{\citenamefont{Schmidt et~al.}(2018)\citenamefont{Schmidt, Avitabile,
  Montbri\'o, and Roxin}}]{SAM+18}
\bibinfo{author}{\bibfnamefont{H.}~\bibnamefont{Schmidt}},
  \bibinfo{author}{\bibfnamefont{D.}~\bibnamefont{Avitabile}},
  \bibinfo{author}{\bibfnamefont{E.}~\bibnamefont{Montbri\'o}},
  \bibnamefont{and} \bibinfo{author}{\bibfnamefont{A.}~\bibnamefont{Roxin}},
  \bibinfo{journal}{PLoS Computational Biology} \textbf{\bibinfo{volume}{14}}
  (\bibinfo{year}{2018}).

\bibitem[{\citenamefont{Ratas and Pyragas}(2018)}]{RP18}
\bibinfo{author}{\bibfnamefont{I.}~\bibnamefont{Ratas}} \bibnamefont{and}
  \bibinfo{author}{\bibfnamefont{K.}~\bibnamefont{Pyragas}},
  \bibinfo{journal}{Phys. Rev. E} \textbf{\bibinfo{volume}{98}},
  \bibinfo{pages}{052224} (\bibinfo{year}{2018}).

\bibitem[{\citenamefont{di~Volo and Torcini}(2018)}]{DT18}
\bibinfo{author}{\bibfnamefont{M.}~\bibnamefont{di~Volo}} \bibnamefont{and}
  \bibinfo{author}{\bibfnamefont{A.}~\bibnamefont{Torcini}},
  \bibinfo{journal}{Phys. Rev. Lett.} \textbf{\bibinfo{volume}{121}},
  \bibinfo{pages}{128301} (\bibinfo{year}{2018}).

\bibitem[{\citenamefont{Dumont and Gutkin}(2019)}]{DG19}
\bibinfo{author}{\bibfnamefont{G.}~\bibnamefont{Dumont}} \bibnamefont{and}
  \bibinfo{author}{\bibfnamefont{B.}~\bibnamefont{Gutkin}},
  \bibinfo{journal}{PLOS Computational Biology} \textbf{\bibinfo{volume}{15}},
  \bibinfo{pages}{1} (\bibinfo{year}{2019}).

\bibitem[{\citenamefont{Akao et~al.}(2019)\citenamefont{Akao, Shirasaka, Jimbo,
  Ermentrout, and Kotani}}]{ASJ+19}
\bibinfo{author}{\bibfnamefont{A.}~\bibnamefont{Akao}},
  \bibinfo{author}{\bibfnamefont{S.}~\bibnamefont{Shirasaka}},
  \bibinfo{author}{\bibfnamefont{Y.}~\bibnamefont{Jimbo}},
  \bibinfo{author}{\bibfnamefont{B.}~\bibnamefont{Ermentrout}},
  \bibnamefont{and} \bibinfo{author}{\bibfnamefont{K.}~\bibnamefont{Kotani}},
  \bibinfo{journal}{arXiv preprint arXiv:1903.12155}  (\bibinfo{year}{2019}).

\bibitem[{\citenamefont{Bi et~al.}(2019)\citenamefont{Bi, Segneri, di~Volo, and
  Torcini}}]{BSD+19}
\bibinfo{author}{\bibfnamefont{H.}~\bibnamefont{Bi}},
  \bibinfo{author}{\bibfnamefont{M.}~\bibnamefont{Segneri}},
  \bibinfo{author}{\bibfnamefont{M.}~\bibnamefont{di~Volo}}, \bibnamefont{and}
  \bibinfo{author}{\bibfnamefont{A.}~\bibnamefont{Torcini}},
  \bibinfo{journal}{bioRxiv}  (\bibinfo{year}{2019}).

\bibitem[{\citenamefont{Coombes and Byrne}(2019)}]{CB19}
\bibinfo{author}{\bibfnamefont{S.}~\bibnamefont{Coombes}} \bibnamefont{and}
  \bibinfo{author}{\bibfnamefont{{\'A}.}~\bibnamefont{Byrne}}, in
  \emph{\bibinfo{booktitle}{Nonlinear Dynamics in Computational Neuroscience}},
  edited by \bibinfo{editor}{\bibfnamefont{F.}~\bibnamefont{Corinto}}
  \bibnamefont{and} \bibinfo{editor}{\bibfnamefont{A.}~\bibnamefont{Torcini}}
  (\bibinfo{publisher}{Springer International Publishing},
  \bibinfo{address}{Cham}, \bibinfo{year}{2019}), pp. \bibinfo{pages}{1--16}.

\bibitem[{\citenamefont{Keeley et~al.}(2019)\citenamefont{Keeley, Byrne,
  Fenton, and Rinzel}}]{KBF+19}
\bibinfo{author}{\bibfnamefont{S.}~\bibnamefont{Keeley}},
  \bibinfo{author}{\bibfnamefont{A.}~\bibnamefont{Byrne}},
  \bibinfo{author}{\bibfnamefont{A.}~\bibnamefont{Fenton}}, \bibnamefont{and}
  \bibinfo{author}{\bibfnamefont{J.}~\bibnamefont{Rinzel}},
  \bibinfo{journal}{Journal of Neurophysiology} \textbf{\bibinfo{volume}{121}},
  \bibinfo{pages}{2181} (\bibinfo{year}{2019}).

\bibitem[{\citenamefont{Byrne et~al.}(2019)\citenamefont{Byrne, Avitabile, and
  Coombes}}]{BAC19}
\bibinfo{author}{\bibfnamefont{A.}~\bibnamefont{Byrne}},
  \bibinfo{author}{\bibfnamefont{D.}~\bibnamefont{Avitabile}},
  \bibnamefont{and} \bibinfo{author}{\bibfnamefont{S.}~\bibnamefont{Coombes}},
  \bibinfo{journal}{Phys. Rev. E} \textbf{\bibinfo{volume}{99}},
  \bibinfo{pages}{012313} (\bibinfo{year}{2019}).

\bibitem[{\citenamefont{Boari et~al.}(2019)\citenamefont{Boari, Uribarri,
  Amador, and Mindlin}}]{BUA+19}
\bibinfo{author}{\bibfnamefont{S.}~\bibnamefont{Boari}},
  \bibinfo{author}{\bibfnamefont{G.}~\bibnamefont{Uribarri}},
  \bibinfo{author}{\bibfnamefont{A.}~\bibnamefont{Amador}}, \bibnamefont{and}
  \bibinfo{author}{\bibfnamefont{G.~B.} \bibnamefont{Mindlin}},
  \bibinfo{journal}{Mathematical and Computational Applications}
  \textbf{\bibinfo{volume}{24}} (\bibinfo{year}{2019}).

\bibitem[{\citenamefont{Laing}(2015)}]{Lai15}
\bibinfo{author}{\bibfnamefont{C.~R.} \bibnamefont{Laing}},
  \bibinfo{journal}{SIAM Journal on Applied Dynamical Systems}
  \textbf{\bibinfo{volume}{14}}, \bibinfo{pages}{1899} (\bibinfo{year}{2015}).

\bibitem[{\citenamefont{Paz\'o and Montbri\'o}(2006)}]{PM06}
\bibinfo{author}{\bibfnamefont{D.}~\bibnamefont{Paz\'o}} \bibnamefont{and}
  \bibinfo{author}{\bibfnamefont{E.}~\bibnamefont{Montbri\'o}},
  \bibinfo{journal}{Phys. Rev. E} \textbf{\bibinfo{volume}{73}},
  \bibinfo{pages}{055202} (\bibinfo{year}{2006}).

\bibitem[{\citenamefont{Kuramoto}(1975)}]{Kur75}
\bibinfo{author}{\bibfnamefont{Y.}~\bibnamefont{Kuramoto}}, in
  \emph{\bibinfo{booktitle}{International Symposium on Mathematical Problems in
  Theoretical Physics}}, edited by
  \bibinfo{editor}{\bibfnamefont{H.}~\bibnamefont{Araki}}
  (\bibinfo{publisher}{Springer}, \bibinfo{address}{Berlin},
  \bibinfo{year}{1975}), vol.~\bibinfo{volume}{39} of
  \emph{\bibinfo{series}{Lecture Notes in Physics}}, pp.
  \bibinfo{pages}{420--422}.

\bibitem[{\citenamefont{Hansel and Mato}(2001)}]{HM01}
\bibinfo{author}{\bibfnamefont{D.}~\bibnamefont{Hansel}} \bibnamefont{and}
  \bibinfo{author}{\bibfnamefont{G.}~\bibnamefont{Mato}},
  \bibinfo{journal}{Phys. Rev. Lett.} \textbf{\bibinfo{volume}{86}},
  \bibinfo{pages}{4175} (\bibinfo{year}{2001}).

\bibitem[{\citenamefont{Hansel and Mato}(2003)}]{HM03}
\bibinfo{author}{\bibfnamefont{D.}~\bibnamefont{Hansel}} \bibnamefont{and}
  \bibinfo{author}{\bibfnamefont{G.}~\bibnamefont{Mato}},
  \bibinfo{journal}{Neural Computation} \textbf{\bibinfo{volume}{15}},
  \bibinfo{pages}{1} (\bibinfo{year}{2003}).

\bibitem[{\citenamefont{Gutkin et~al.}(2001)\citenamefont{Gutkin, Laing, Colby,
  Chow, and Ermentrout}}]{GLC+01}
\bibinfo{author}{\bibfnamefont{B.~S.} \bibnamefont{Gutkin}},
  \bibinfo{author}{\bibfnamefont{C.~R.} \bibnamefont{Laing}},
  \bibinfo{author}{\bibfnamefont{C.~L.} \bibnamefont{Colby}},
  \bibinfo{author}{\bibfnamefont{C.~C.} \bibnamefont{Chow}}, \bibnamefont{and}
  \bibinfo{author}{\bibfnamefont{G.~B.} \bibnamefont{Ermentrout}},
  \bibinfo{journal}{Journal of Computational Neuroscience}
  \textbf{\bibinfo{volume}{11}}, \bibinfo{pages}{121} (\bibinfo{year}{2001}).

\bibitem[{\citenamefont{Laing and Chow}(2001)}]{LC01}
\bibinfo{author}{\bibfnamefont{C.~R.} \bibnamefont{Laing}} \bibnamefont{and}
  \bibinfo{author}{\bibfnamefont{C.~C.} \bibnamefont{Chow}},
  \bibinfo{journal}{Neural Computation} \textbf{\bibinfo{volume}{13}},
  \bibinfo{pages}{1473} (\bibinfo{year}{2001}).

\bibitem[{\citenamefont{Kanamaru and Sekine}(2003)}]{KS03}
\bibinfo{author}{\bibfnamefont{T.}~\bibnamefont{Kanamaru}} \bibnamefont{and}
  \bibinfo{author}{\bibfnamefont{M.}~\bibnamefont{Sekine}},
  \bibinfo{journal}{Phys. Rev. E} \textbf{\bibinfo{volume}{67}},
  \bibinfo{pages}{031916} (\bibinfo{year}{2003}).

\bibitem[{\citenamefont{Sakaguchi et~al.}(1988)\citenamefont{Sakaguchi,
  Shinomoto, and Kuramoto}}]{SSK88}
\bibinfo{author}{\bibfnamefont{H.}~\bibnamefont{Sakaguchi}},
  \bibinfo{author}{\bibfnamefont{S.}~\bibnamefont{Shinomoto}},
  \bibnamefont{and} \bibinfo{author}{\bibfnamefont{Y.}~\bibnamefont{Kuramoto}},
  \bibinfo{journal}{Progress of Theoretical Physics}
  \textbf{\bibinfo{volume}{79}}, \bibinfo{pages}{600} (\bibinfo{year}{1988}).

\bibitem[{\citenamefont{Zaks et~al.}(2003)\citenamefont{Zaks, Neiman, Feistel,
  and Schimansky-Geier}}]{ZNF+03}
\bibinfo{author}{\bibfnamefont{M.~A.} \bibnamefont{Zaks}},
  \bibinfo{author}{\bibfnamefont{A.~B.} \bibnamefont{Neiman}},
  \bibinfo{author}{\bibfnamefont{S.}~\bibnamefont{Feistel}}, \bibnamefont{and}
  \bibinfo{author}{\bibfnamefont{L.}~\bibnamefont{Schimansky-Geier}},
  \bibinfo{journal}{Phys. Rev. E} \textbf{\bibinfo{volume}{68}},
  \bibinfo{pages}{066206} (\bibinfo{year}{2003}).

\bibitem[{\citenamefont{M.Childs and Strogatz}(2008)}]{CS08}
\bibinfo{author}{\bibfnamefont{L.}~\bibnamefont{M.Childs}} \bibnamefont{and}
  \bibinfo{author}{\bibfnamefont{S.~H.} \bibnamefont{Strogatz}},
  \bibinfo{journal}{Chaos} \textbf{\bibinfo{volume}{18}}, \bibinfo{eid}{043128}
  (\bibinfo{year}{2008}).

\bibitem[{\citenamefont{Lafuerza et~al.}(2010)\citenamefont{Lafuerza, Colet,
  and Toral}}]{LCT10}
\bibinfo{author}{\bibfnamefont{L.~F.} \bibnamefont{Lafuerza}},
  \bibinfo{author}{\bibfnamefont{P.}~\bibnamefont{Colet}}, \bibnamefont{and}
  \bibinfo{author}{\bibfnamefont{R.}~\bibnamefont{Toral}},
  \bibinfo{journal}{Phys. Rev. Lett.} \textbf{\bibinfo{volume}{105}},
  \bibinfo{pages}{084101} (\bibinfo{year}{2010}).

\bibitem[{\citenamefont{Daido and Nakanishi}(2004)}]{DN04}
\bibinfo{author}{\bibfnamefont{H.}~\bibnamefont{Daido}} \bibnamefont{and}
  \bibinfo{author}{\bibfnamefont{K.}~\bibnamefont{Nakanishi}},
  \bibinfo{journal}{Phys. Rev. Lett.} \textbf{\bibinfo{volume}{93}},
  \bibinfo{pages}{104101} (\bibinfo{year}{2004}).

\bibitem[{\citenamefont{Kuramoto}(1984)}]{Kur84}
\bibinfo{author}{\bibfnamefont{Y.}~\bibnamefont{Kuramoto}},
  \emph{\bibinfo{title}{Chemical Oscillations, Waves, and Turbulence}}
  (\bibinfo{publisher}{{S}pringer-{V}erlag}, \bibinfo{address}{Berlin},
  \bibinfo{year}{1984}).

\end{thebibliography}

 \makeatletter
\setcounter{equation}{0}
\renewcommand{\theequation}{A\arabic{equation}}
 \makeatother

\section*{Appendix A: Derivation of the Firing Rate Equations}

In the thermodynamic limit, $N\to \infty$, we drop the indices for the individual neuronal
dynamics Eq.~\eqref{qif}, 
and denote $\rho(V \vert \eta, t) d V$ as the fraction of neurons with 
membrane potentials between $V$ and $V+dV$, and parameter $\eta$ at time $t$.
Accordingly, the parameter $\eta$ becomes a continuous random variable
that is distributed according to a probability distribution function, which here is considered 
to be a Lorentzian $L_{\Delta,\bar\eta}(\eta)$
of half-width $\Delta$ and centered at $\bar \eta$, see Eq.~\eqref{lorentzian2}.
The conservation of the number of neurons leads to the continuity equation
\begin{equation}
\tau\partial_t \rho + \partial_V\left[  ( V^2+\eta + g(v-V) +J\tau  r ) \rho \right]=0 ,
\label{cont}
\end{equation}
where we explicitly included the velocity given by the continuous equivalent of Eq.~\eqref{qif}.
We also defined the mean value of the membrane potential as 
\begin{equation}
v(t)=\int_{-\infty}^{\infty}  \int_{-\infty}^{\infty} \rho(V| \eta,t)~V L_{\Delta,\bar\eta}(\eta)~dVd\eta.
\label{v}
\end{equation}
Next, we consider the family of conditional density functions~\cite{MPR15}
\begin{equation}
\rho(V \vert \eta, t)= \frac{1}{\pi}\frac{x(\eta,t)}{\left[V-y(\eta,t)\right]^2
+x(\eta,t)^2},
\label{la}
\end{equation}
which are Lorentzian functions with time-dependent half-width
$x(\eta,t)$, centered at $y(\eta,t)$.
Substituting \eqref{la} into the continuity equation
\eqref{cont}, we find that, for each value of $\eta$,
variables $x$ and $y$ must obey two coupled equations,
\begin{subequations}
\label{xy-eqs}
\begin{eqnarray}
\tau \dot x (\eta,t)&=& 2 x (\eta,t) y (\eta,t) - g x (\eta,t), \label{xeta}\\
\tau \dot y (\eta,t)&=& \eta -  x (\eta,t)^2 + y (\eta,t)^2 \label{yeta}\\
&& \hspace{.85cm} - g \big[ y(\eta,t) - v \big] + J\tau r, \nonumber
\end{eqnarray}
\end{subequations}
that can be written in complex form as
\begin{equation}
\tau \partial_t w(\eta,t)=  i\big[  \eta -w(\eta,t)^2+J \tau r \big]
+ g \big[i v- w (\eta,t) \big]
\label{w}
\end{equation}
where $w(\eta,t)\equiv x(\eta,t) +i y(\eta,t)$.
For a particular value of $\eta$, the firing rate $r$ of the population of QIF
neurons is related to the width $x$ of the Lorentzian ansatz~\eqref{la}. 
Specifically, the firing rate $r(\eta,t)$ for each $\eta$ value at time $t$ 
is the probability flux at infinity: 
$r(\eta,t)= \rho(V\to \infty |\eta,t) \dot V (V\to \infty|\eta,t)$, 
which yields the identity
\begin{equation}
x(\eta,t)=\pi \tau  r(\eta,t).
\label{x}
\end{equation}
Hence, integrating this quantity over the distributions of 
currents Eq.~\eqref{lorentzian2} provides the mean firing rate
\begin{equation}
r(t)=\frac{1}{\tau\pi}\int_{-\infty}^{\infty} x(\eta,t) L_{\Delta,\bar\eta}(\eta)d\eta.
\label{r}
\end{equation}
Likewise, we can link the center $y(\eta,t)$ of the Lorentzian ansatz
Eq.~\eqref{la} with the mean of the (conditional) membrane potential via
\begin{equation}
y(\eta,t)= \text{p.v.} \int_{-\infty}^{\infty} \rho(V \vert \eta,t) VdV.
\label{yPV}
\end{equation}
Note that the Lorentzian distribution does not have finite moments so that 
the integral in Eq.~\eqref{yPV} needs to be taken as the Cauchy 
principal value (i.e. $ \text{p.v.} \int_{-\infty}^\infty \rho VdV =
\lim_{R \to \infty} \int_{-R}^{R} \rho VdV $).
Then, Eq.~\eqref{v} becomes
\begin{equation}
v(t)= \int_{-\infty}^{\infty}y(\eta,t) L_{\Delta,\bar\eta}(\eta)d\eta.
\label{vnew}
\end{equation}

The integrals in (\ref{r},\ref{vnew}) can be evaluated closing the integral contour
in the complex $\eta$-plane and using Cauchy's residue theorem. The integrals must
however be performed carefully, so that the variable $x(\eta,t)$ remains non-negative.
To make the analytic continuation of $w(\eta,t)$ from real to complex-valued $\eta$,
we define $\eta \equiv \eta_r+i\eta_i$.
This continuation is possible into the lower half-plane $\eta_i<0$, since 
this guarantees the half-width $x(\eta,t)$ to remain non-negative:
$\partial_t x(\eta,t)=- \eta_i >0$, at $x=0$. Therefore, we perform contour 
integration in Eq.~\eqref{r} and Eq.~\eqref{vnew} along the
arc $|\eta| e^{i\vartheta}$ with $|\eta|\to\infty$ and $\vartheta\in(-\pi,0)$.
This contour encloses one pole of the Lorentzian distribution Eq.~\eqref{lorentzian2}.
Then, we find that the firing rate and the mean membrane potential depend only 
on the value of $w$ at
the pole of $L_{\Delta,\bar\eta}(\eta)$ in the lower half $\eta$-plane:
$$\pi \tau r(t)+i v(t) = w(\bar\eta-i\Delta,t),$$
As a result, we only need to evaluate Eq.~\eqref{w} at $\eta=\bar\eta-i\Delta$, 
and obtain a system of FRE composed of two ordinary differential equations as given in Eq.~\eqref{fre},
\begin{subequations}
\begin{eqnarray}
\tau \dot r &=& \frac{\Delta}{\tau \pi} + 2  r v - g r, \nonumber \\
\tau \dot v &=&   v^2 +   \bar \eta - (\pi \tau r)^2 + J \tau r  \nonumber , 
\end{eqnarray}
\end{subequations}
in terms of the population-mean firing rate $r$ and the population-mean membrane potential
 $v$. Multiplying the Lorentzian ansatz Eq.~\eqref{la} by $L_{\Delta,\bar\eta}(\eta)$ and integrating over $\eta$, we finally obtain the total density of neurons Eq.~\eqref{la0} as
\begin{equation}
\rho(V, t)= \frac{1}{\pi}\frac{ \pi \tau r(t)}{[V-v(t)]^2+\pi^2 \tau^2 r(t)^2}, 
\nonumber 
\end{equation}
where we again applied Cauchy's residue theorem by using that the ansatz Eq.~\eqref{la} 
is analytic in the lower $\eta$-complex plane. Hence, the total 
density of the population of QIF neurons is a Lorentzian distribution centered at
$v(t)$ and half-width $\pi \tau r(t)$, which evolves according to the FRE~\eqref{fre}.


 \makeatletter
\setcounter{equation}{0}
\renewcommand{\theequation}{B\arabic{equation}}
 \makeatother

\section*{Appendix B: Connection between the FRE in~\cite{Lai15} and Eq.~(\textbf{\ref{fre}})}

The derivation of the FRE~\eqref{fre} 
is exact in the thermodynamic limit, and does not rely on any approximation.
Here we show that the Eqs.~(2.35\&2.36) in~\cite{Lai15} 
reduce to our Eq.~\eqref{fre} after adopting a limit  
in which the derivation performed in~\cite{Lai15} becomes exact. 

In contrast to our Eq.~\eqref{freb}, note that Eq.~(2.36) in~\cite{Lai15} contains a diffusive term,
\begin{equation}
g[Q(t)-v(t)],
\label{diff}
\end{equation}
where the function $Q(t)$ is defined as
\begin{equation}
Q(t)=\frac{i}{2}\sum_{m=1}^{\infty} \frac{\rho^{m+1}-\rho^{m-1}}{\rho +1+\epsilon}
\left[ z^m -  \bar z^m\right]
\label{Q}
\end{equation}
with $0<\epsilon\ll 1$, and $\rho = \sqrt{2\epsilon+\epsilon^2}-1-\epsilon$.
The variable $z$ in Eq.~\eqref{Q} is the complex Kuramoto order parameter 
(the bar denotes complex conjugation), which is 
related to the variables $r$ and $v$ in the FRE~\eqref{fre} 
via the change of variables~\cite{MPR15}
\begin{equation}
\pi r+iv=\frac{1-\bar z}{1+\bar z}.
\label{map}
\end{equation}
The parameter $\epsilon$, defined in Eq.~(2.7) in~\cite{Lai15}, 
was used to approximate the mean voltage $v$, see also~\cite{Erm06}. 
In the limit $\epsilon\to 0$ this approximation becomes exact, 
but this limit was not considered in~\cite{Lai15}. 
In consequence, to use the Eqs.~(2.35\&2.36) in \cite{Lai15},
the infinite series Eq.~\eqref{Q} was truncated after 100 terms, and 
the bifurcation analysis of the mean-field model could only be 
performed numerically.

Using the geometric series formula ($|z|<1$) and 
the transformation of variables Eq.~\eqref{map}, we find 
\begin{equation}
\lim_{\epsilon\to 0} Q(t)=i \sum_{m=1}^{\infty}  (-1)^m \left[ z^m - \bar z^m 
\right]=\frac{2\text{Im}(z)}{(1+z)(1+\bar z)} =v.
\nonumber
\end{equation}
Hence we have showed that the 
diffusive term Eq.~\eqref{diff} 
identically vanishes when the mean-field reduction 
becomes exact (i.e.~in the limit $\epsilon \to 0$), and the FRE in~\cite{Lai15} 
reduce to Eqs.~\eqref{fre}.

\setcounter{equation}{0}
\renewcommand{\theequation}{C\arabic{equation}}
 \makeatother


\section*{Appendix C: Bifurcation analysis of the Firing Rate Equations}

The FRE~\eqref{fre} have five free parameters. The number of effective
parameters can be reduced to three
through non-dimensionalization, defining
$$\tilde \eta= \bar \eta/\Delta,~\tilde g= g/\sqrt{\Delta},
~\tilde J= J/(\pi \sqrt{\Delta}),$$
and rescaling variables as
$$\tilde r= \tau \pi r /\sqrt{\Delta} ,~
\tilde v= v/ \sqrt{\Delta},~
\tilde t= \sqrt \Delta t/\tau.$$ 
Then, the firing rate model becomes
\begin{subequations}
\label{fre0}
\begin{eqnarray}
\frac{d \tilde r}{d\tilde t} &=& 1 + 2  \tilde r  \tilde v -  \tilde g  \tilde r, \label{fre0R}\\
\frac{d \tilde v}{d\tilde t} &=&   \tilde v^2 +\tilde \eta- \tilde  r^2 + 
 \tilde J  \tilde r,
\label{fre0V}
\end{eqnarray}
\end{subequations}
The fixed points $(\tilde r_*,\tilde v_*)$ of Eq.~\eqref{fre0} satisfy
\begin{equation}
\tilde v_*=\frac{\tilde g }{2}-\frac{1}{2\tilde r_*}.
\label{vfp}
\end{equation}
Linearization about the fixed points Eq.~\eqref{vfp} gives the eigenvalues
\begin{equation}
\lambda_\pm= \frac{1}{ 2} \left( 4 \tilde v_* -
\tilde g \pm \sqrt{ \tilde g^2 +  8   \tilde r_* ( \tilde J- 2   \tilde r_* ) } \right).
\label{lambda}
\end{equation}
For networks with only chemical synapses (i.e.~$g=0$), the real part of the eigenvalues
remains always negative (since $v_*<0$), and a Hopf bifurcation is not
possible. However, chemical coupling has a direct influence on the
real part of the eigenvalues Eq.~\eqref{lambda}, and may produce oscillatory    
instabilities if the argument of the square root is a real number.

\subsection{Hopf boundaries and Takens-Bogdanov point}

The Hopf boundaries can be obtained when imposing Re($\lambda_\pm)=0$
in Eq.~\eqref{lambda}, which gives
$\tilde g=4 \tilde v_*$. Then, using Eq.~\eqref{vfp}, we find 
\begin{equation}
\tilde g_H=2/\tilde r_*.
\label{gH2}
\end{equation}
Substituting Eq.~\eqref{gH2} in the $v$-fixed point equation Eq.~\eqref{fre0V},
and solving for $\tilde \eta$, we obtain
\begin{equation}
\tilde \eta_H=  r_*^2-\tilde J \tilde r_* - \frac{1 }{4 \tilde  r_*^2}.
\label{etaH2}
\end{equation}
Solving Eq.~\eqref{gH2} for $\tilde r_*$ and substituting it into
Eq.~\eqref{etaH2} we obtain the Hopf boundaries in explicit form
\begin{equation}
\tilde \eta_H=-\frac{2 \tilde J}{ \tilde g} + \frac{4}{\tilde g^2} -\frac{\tilde g^2 }{16}.
\label{Hopf}
\end{equation}
The frequency of the oscillations
is given by the imaginary part of the eigenvalues Eq.~\eqref{lambda}
at criticality that, using the fixed points of Eqs.~\eqref{fre0} and 
Eq.~\eqref{gH2}, reduces to the explicit formula
\begin{equation}
\tilde f_H= \frac{1}{\pi}\sqrt{ \tilde \eta +\frac{\tilde J}{\tilde g}}.
\label{fH2}
\end{equation}
The frequency becomes zero at a Takens-Bogdanov (TB) point, when 
$\tilde \eta=-\tilde J/\tilde g$. Inserting this condition
into Eq.~\eqref{Hopf} we obtain the coordinates of the TB point
\begin{equation}
\left(\tilde \eta,\tilde  J \right)_{\text{\scriptsize{TB}}} =
\left( \frac{\tilde g^2 }{16}- \frac{4}{\tilde g^2} ,
\frac{4}{\tilde  g} -\frac{\tilde g^3 }{16 }\right),
\label{TB}
\end{equation}
see also Fig.~\ref{Fig6}. For $\tilde J=0$, the TB point is located at
\begin{equation}
\left(\tilde \eta,\tilde  g \right)_{\text{\scriptsize{TB}}} =
\left(0,2\sqrt{2} \right)
\label{TB0}
\end{equation}
in the phase diagram Fig.~\ref{Fig5}.

\subsection{Saddle-node boundaries}

The boundaries of the saddle-node bifurcations are obtained by setting
$\lambda_\pm=0$ in Eq.~\eqref{lambda}, using Eq.~\eqref{vfp}, and solving for $\tilde g$:
\begin{equation}
\tilde g_{sn}=\frac{1}{ \tilde r_*}-2  \tilde J \tilde r_*^2+4  \tilde r_*^3 \  .
\label{gSN1}
\end{equation}
Substituting~\eqref{gSN1} in the $v$-fixed point Eq.~\eqref{fre0V}, and solving for
$\tilde \eta$, we obtain
\begin{equation}
\tilde \eta_{sn}= \tilde r_*^2 - 4 \tilde r^6_*+
\tilde J  \tilde  r_* \left(   4 \tilde r_*^4 - J \tilde r_*^3-1\right) .
\label{etaSN1}
\end{equation}
The saddle node boundaries are plotted in the $(\tilde \eta,\tilde g)$
phase diagram in Fig.~\ref{Fig5}. The same boundaries can be represented in the
$(\tilde \eta,\tilde J)$ phase diagram when solving Eq.~\eqref{gSN1}
for $\tilde J$  
\begin{equation}
\tilde J_{sn}=\frac{1}{2 \tilde r_*^3 }- \frac{\tilde g}{2\tilde r_*^2}+2  \tilde r_*
\label{JSN2}
\end{equation}
and replacing $\tilde J$ by Eq.~\eqref{JSN2} in Eq.~\eqref{etaSN1} gives
\begin{equation}
\tilde \eta_{sn}=\frac{\tilde g}{\tilde r_*}-
\frac{\tilde g^2}{4} -\frac{3}{4 \tilde r_*^2} -r_*^2 .
\nonumber
\end{equation}
These saddle-node
boundaries are shown in black for different values of $\tilde g$ in
Fig.~\ref{Fig6}.

\subsection{Focus-Node boundaries}
The boundaries in the phase diagram 
Fig.~\ref{Fig6} in which the stable asynchronous state
changes from Focus to Node can be obtained in parametric form 
equating the square root in Eq.~\eqref{lambda} to zero. 
This gives 
\begin{equation}
\tilde J_{FN}=\frac{16\tilde r_*^2- \tilde g^2}{8\tilde r_*}.
\nonumber
\end{equation}
Substituting $\tilde J_{FN}$ into the $v$-fixed point Eq.~\eqref{fre0V}, and using Eq.~\eqref{vfp}
we find 
\begin{equation}
\tilde \eta_{FN}=\frac{1}{8\tilde r_*^2} \left(-2+4 \tilde g \tilde r_*- \tilde g^2 \tilde r_*^2-
8\tilde r_*^4 \right).
\nonumber
\end{equation}
For networks without chemical coupling, $\tilde J=0$, the Focus-Node boundary can be
obtained in explicit form   
\begin{equation}
\tilde \eta_{FN}=2-\frac{4}{\tilde g^2}-\frac{3\tilde g^2}{16}.
\nonumber
\end{equation}
This is the dashed boundary depicted in Fig.~\ref{Fig5}.

\setcounter{equation}{0}
\renewcommand{\theequation}{D\arabic{equation}}
 \makeatother


\section*{Appendix D: Small-amplitude equation 
near the Hopf bifurcation}
 
In this Appendix we derive the small amplitude equation 
near the Hopf bifurcation, and show that the Hopf bifurcation is 
always supercritical. The derivation is performed using 
multiple-scales analysis, see e.g.~\cite{Kur84}. 
We first expand the solution    
\begin{equation}
\binom{\tilde r}{\tilde v}=
\binom{r_0}{v_0}+ 
\epsilon \binom{r_1}{v_1}+\epsilon^2 \binom{r_2}{v_2}  
+ \dots
\label{expansion0}
\end{equation}
in powers of a small parameter $\epsilon \ll 1$, about a fixed point  
$(r_0,v_0)$ of Eqs.~\eqref{fre0} at the Hopf bifurcation. 
In addition, we introduce the deviation from the Hopf bifurcation 
Eq.~\eqref{Hopf} of parameter $\tilde \eta$ as
\begin{equation}
\tilde \eta- \tilde \eta_H=\chi \epsilon^2,
\label{deviation}
\end{equation}
where $\chi$ determines the sign of the deviation. Finally, we 
define the slow time 
\begin{equation}
T=\epsilon^2 t.
\end{equation}
Then, the time differentiation is transformed as 
\begin{equation}
\frac{d}{dt}\to \partial_t+\epsilon^2 \partial_T.
\label{dt}
\end{equation}
Plugging Eqs.~(\ref{expansion0},\ref{deviation},\ref{dt})
into Eq.~\eqref{fre0} gives 
\begin{align}
&\left[\partial_t + \epsilon^2 \partial_T     
 -L_0 \right]
\left[ 
\epsilon \binom{r_1}{v_1}+\epsilon^2 \binom{r_2}{v_2}  
+ \dots \right] 
- \epsilon^2 \binom{0}{\chi}= \nonumber \\ 
&\hspace*{4cm}=\epsilon^2 N_2+\epsilon^3 N_3+\dots,
\label{expansion}
\end{align}
where 
\begin{equation}  
  L_0=\begin{pmatrix} 
    2 v_0- \tilde g  & 2 r_0 \\ 
    \tilde J-2 r_0  & 2 v_0    
  \end{pmatrix} .
  \label{L00} 
\end{equation} 
and 
\begin{equation}  
  N_2=\binom{2  r_1 v_1}{ v_1^2 - r_1^2},\quad
  N_3=\binom{2r_1 v_2 + 2r_2 v_1}
	{2v_1  v_2-2 r_1 r_2}.
  \label{N2} 
\end{equation} 

\subsection*{Critical eigenvectors} 

Using Eqs.~(\ref{lambda},\ref{gH2}), we define the critical frequency as 
\begin{equation}
\omega_0=\frac{1}{2\tilde g}\sqrt{64-\tilde g^4 -16 \tilde J \tilde g},
\label{omega}
\end{equation}
so that the matrix Eq.~\eqref{L00} can be written as 
\begin{equation}  
  L_0=\begin{pmatrix} 
    -\tilde g/2  & 4/\tilde g  \\ 
    -\tilde g/16(\tilde g^2 +4\omega_0^2)  & \tilde g/2 
  \end{pmatrix} .
  \label{L0} 
\end{equation} 
The right-eigenvector of Eq.~\eqref{L0} is   
\begin{equation} 
  \mathbf{u}_R=\binom{4/\tilde g}{ \tilde g/2+ i \omega_0}. 
  \label{uR} 
\end{equation}  
Imposing the condition $\mathbf{u}_L \mathbf{u}_{R}=1$, the 
left-eigenvector of $L_0$ is 
\begin{equation} 
  \mathbf{u}_L=\frac{1}{2\omega_0}\left(\frac{2\tilde g \omega_0+i \tilde g^2 }{8},-i\right).
  \label{uL} 
\end{equation}  

\subsection*{Analysis of multiple scales} 

At order $\epsilon$, Eq.~\eqref{expansion} is 
\begin{equation} 
\binom{\dot r_1}{\dot v_1}-L_0 \binom{r_1}{v_1}=\binom{0}{0}.  
\end{equation}  
This system of differential equations has a general solution 
\begin{equation} 
\binom{r_1}{v_1}= A e^{i \omega_0  \tilde t} \mathbf{u}_R + \text{c.c}, 
\label{nSol} 
\end{equation}  
which is the so-called neutral solution.

At order $\epsilon^2$, Eq.~\eqref{expansion} is 
\begin{equation} 
\binom{\dot r_2}{\dot v_2}- L_0 \binom{r_2}{v_2}- \binom{0}{\chi}= N_2. 
\label{O2} 
\end{equation}  
Substituting the neutral solution Eq.~\eqref{nSol} into 
Eq.~\eqref{N2} we find 
\begin{eqnarray} 
N_2&=& \binom{8}{(\tilde g^4+4\tilde g^2 \omega_0^2-64)/(2\tilde g^2)}|A|^2 
\nonumber\\&+&
\binom{4+8 i \omega_0/\tilde g  }{ \tilde g^2/4 + i \tilde g \omega_0- \omega_0^2 -16/\tilde g^2} A^2 e^{2i\omega_0 \tilde t} \nonumber \\&+&\text{c.c.} \nonumber
\end{eqnarray}  
Next we use the following ansatz  
\begin{equation} 
\binom{r_2}{v_2}=\binom{r_{20}}{v_{20}}+\binom{r_{22}}{v_{22}} e^{2i\omega_0 \tilde t}  +\text{c.c.}
\label{u2}
\end{equation}  
and substitute it into Eq.~\eqref{O2}. We find 
\begin{eqnarray}
r_{20}&=& \frac{2}{\tilde g^3 \omega_0^2} \left[ 2 \tilde g^2 \chi -( 64+\tilde g^4 -4
\tilde g^2 \omega_0^2 ) |A|^2\right], \nonumber\\
v_{20}&=& \frac{1}{4 \tilde g \omega_0^2} \left[2 \tilde g^2  \chi -(64 +\tilde g^4 +4
\tilde g^2\omega_0^2 ) |A|^2\right], \nonumber\\
r_{22}&=& \frac{A^2}{3 \tilde g^3 \omega_0^2} \left[ 64+\tilde g^2 (\tilde g+ 2 i \omega_0)
(\tilde g-10 i \omega_0)  \right], \nonumber\\
v_{22}&=& \frac{A^2}{24 \tilde g^2 \omega_0^2} \left[\tilde g (64+\tilde g( \tilde g+ 2 i \omega_0)^2 (\tilde g-8 i \omega_0)) +256 i \omega_0  \right]. \nonumber
\end{eqnarray}\\
%
Substituting Eqs.~(\ref{nSol},\ref{u2}) into the cubic term $N_3$ in Eq.~\eqref{N2}, we find
\begin{widetext}
\begin{equation} 
N_3=\binom{(\tilde g+2i\omega_0)A r_{20} +\tfrac{8}{\tilde g} (A v_{20}+A^*v_{22} )
+(\tilde g-2i\omega_0) A^*r_{22} }
{(\tilde g+2i\omega_0)A v_{20} -\tfrac{8}{\tilde g} (A r_{20}+A^*r_{22} )
+(\tilde g-2i\omega_0) A^*v_{22} }e^{i\omega_0 \tilde t}+\text{c.c.}  +
\binom{\tfrac{8}{\tilde g}v_{22} + (\tilde g+2 i \omega_0 )r_{22} }
{-\tfrac{8}{\tilde g}r_{22} + (\tilde g+2 i \omega_0 )v_{22}} A e^{3i\omega_0 \tilde t}+\text{c.c.}
\label{N3s}
\end{equation}  
\end{widetext}
The solvability condition at order $\epsilon^3$ is 
\begin{equation} 
\int_0^{2\pi/\omega_0}  \mathbf{u}_L \left[\partial_T \binom{r_1}{v_1}- N_3
\right]e^{-i \omega_0 \tilde t}~d \tilde t=0. 
\label{solvability} 
\end{equation}  
Substituting Eqs.~(\ref{uL},\ref{nSol},\ref{N3s}) into Eq.~\eqref{solvability},
we find the amplitude equation 
\begin{equation} 
\partial_T A=(a+ib) \chi A - (c+id)A|A|^2,
\label{AmpEq}
\end{equation}  
with the coefficients
\begin{eqnarray}
a&=& \frac{4 \tilde g^3  }{64-\tilde g^4 -16\tilde g \tilde J},\nonumber\\
b&=& \frac{32 \tilde g (8-\tilde g \tilde J ) }{(64-\tilde g^4 -16\tilde g J)^{3/2}},
\nonumber\\
c&=& \frac{16 \tilde g (8-\tilde g \tilde J)  }{64-\tilde g^4 -16\tilde g J},\nonumber\\
d&=& \frac{16}{3}~ \frac{  3 (64-\tilde g^4) \tilde J-8 \tilde g \tilde J + 40 \tilde g^3}
	{(64-\tilde g^4 -16\tilde g J)^{3/2}},\nonumber
\end{eqnarray}
where we used Eq.~\eqref{omega} to express $a - d$ in terms of $\tilde J$ and $\tilde g$.
Defining the amplitude $R$ and the phase $\Psi$ via 
$$A=R e^{i\psi},$$ one may alternatively
write  Eq.~\eqref{AmpEq} as 
\begin{eqnarray}
R'&=& \chi a R - c R^3,\nonumber\\
\psi' &=& \chi b - d R^2.\nonumber
\end{eqnarray}
where primes refer to differentiation with respect to $T$. 
An oscillatory solution with amplitude $R=R_s$ and 
phase $\psi=\omega T + \psi_0$, with 
\begin{equation} 
R_s=\sqrt{\frac{a}{|c|}},\quad \omega=\chi b - dR_s^2, 
\nonumber
\end{equation}  
appears in the supercritical  ($\chi>0$) region for $c>0$, and in the 
subcritical region for $c<0$. 

Remarkably, the coefficient $c$ is always positive and hence 
the Hopf bifurcation is always supercritical.
This can be seen 
as follows: First, note that the denominator of $c$ remains always 
positive along the Hopf boundary Eq.~\eqref{Hopf}, and becomes zero at the 
Takens-Bogdanov point Eq.~\eqref{TB}. Second, note that 
the numerator of $c$ is positive for $\tilde J=0$ and may potentially 
change sign at $\tilde J_c=8/\tilde g$. However, this change of sign 
always occurs after the TB point 
($\tilde J_c> \tilde J_\text{{\scriptsize{TB}}}$) 
in which the Hopf bifurcation ends, see Eq.~\eqref{TB}.

Finally, the approximate 
solution in terms of the original variables reads 
\begin{equation} 
\binom{r}{v} \approx \binom{r_0}{v_0}+ \epsilon \binom{r_1}{v_1}=
\binom{r_0}{v_0}+\epsilon  R_s \mathbf{u}_R 
 e^{i (\omega_0+ \epsilon^2 \omega) \tilde t} +\text{c.c.} \nonumber,
\end{equation}      
which describes an oscillatory motion in the critical eigenplane, with a 
small amplitude firing rate (for $\tilde \eta \geq \tilde \eta_H$)
\begin{equation} 
r_A=\epsilon R_s \frac{g}{4}=2 \sqrt{\frac{\tilde \eta-\tilde \eta_H}{8-\tilde g\tilde J}}.
\end{equation}  

\end{document}